\documentclass[a4paper,onecolumn,11pt,unpublished]{quantumarticle}

\pdfoutput=1

\usepackage[numbers,sort,compress]{natbib}

\usepackage[utf8]{inputenc}
\usepackage{amsmath}
\usepackage{amssymb}
\usepackage{amsthm}
\usepackage{mathtools}
\usepackage{physics}
\usepackage{graphicx}
\usepackage{xcolor}
\usepackage{hyperref}
\usepackage{xspace}

\usepackage[group-separator={,}, group-minimum-digits=3]{siunitx} % for comma in numbers
\theoremstyle{definition}
\newtheorem{definition}{Definition}

\theoremstyle{definition}
\newtheorem{strategy}{Strategy}

\newcommand{\sizeNormFrac}[1]{Q_\mathrm{size}^{(#1)}}
\newcommand{\llrNormFrac}[1]{Q_\mathrm{LLR}^{(#1)}}
\newcommand{\plog}{p_\mathrm{log}}
\newcommand{\pabort}{p_\mathrm{abort}}
\newcommand{\avgTimeCostAcc}{\bar{T}_\mathrm{accepted}}
\newcommand{\pabortround}{p_\mathrm{abort}^\mathrm{round}}

\newcommand{\subfig}[1]{\textbf{(#1)}}
\newcommand{\lgz}{\bar{Z}}
\newcommand{\lgx}{\bar{X}}

\newcommand{\image}{\operatorname{IMAGE}}

\begin{document}

\title{Efficient Post-Selection for General Quantum LDPC Codes}

\author{Seok-Hyung Lee}
\affiliation{Centre for Engineered Quantum Systems, School of Physics, The University of Sydney, Sydney, New South Wales 2006, Australia}
\affiliation{Department of Quantum Information Engineering, Sungkyunkwan University, Suwon 16419, Republic of Korea}
\email{seokhyunglee@skku.edu}
\thanks{corresponding author}

\author{Lucas H. English}
\affiliation{Centre for Engineered Quantum Systems, School of Physics, The University of Sydney, Sydney, New South Wales 2006, Australia}

\author{Stephen D. Bartlett}
\email{stephen.bartlett@sydney.edu.au}
\affiliation{Centre for Engineered Quantum Systems, School of Physics, The University of Sydney, Sydney, New South Wales 2006, Australia}
\thanks{corresponding author}
 
\begin{abstract}
Post-selection strategies that discard low-confidence computational results can significantly improve the effective fidelity of quantum error correction at the cost of reduced acceptance rates, which can be particularly useful for offline resource state generation and other moderate-depth fault-tolerant circuits.
Prior work has primarily relied on the ``logical gap'' metric with the minimum-weight perfect matching decoder, but this approach faces fundamental limitations including computational overhead that scales exponentially with the number of logical qubits and poor generalizability to arbitrary codes beyond surface codes.
We develop post-selection strategies based on computationally efficient heuristic confidence metrics that leverage error cluster statistics (specifically, aggregated cluster sizes and log-likelihood ratios) from clustering-based decoders, which are applicable to arbitrary quantum low-density parity check (QLDPC) codes.
We validate our method through extensive numerical simulations on surface codes, bivariate bicycle codes, and hypergraph product codes, demonstrating orders of magnitude reductions in logical error rates with moderate abort rates.
For instance, applying our strategy to the $[[144, 12, 12]]$ bivariate bicycle code achieves approximately three orders of magnitude reduction in the logical error rate with an abort rate of only 1\% (19\%) at a physical error rate of 0.1\% (0.3\%).
Additionally, we integrate our approach with the sliding-window framework for real-time decoding, featuring early mid-circuit abort decisions that eliminate unnecessary overheads.
Notably, its performance matches or even surpasses the original strategy for global decoding, while exhibiting favorable scaling in the number of rounds.
Our approach provides a practical foundation for efficient post-selection in fault-tolerant quantum computing with QLDPC codes.
\end{abstract}

\maketitle

\section{Introduction}
\label{sec:introduction}

Quantum computing promises exponential speedups for certain computational problems, but realizing this potential requires overcoming the fundamental challenge of quantum decoherence and operational errors \cite{terhal2015quantum}. 
Quantum error correction (QEC) provides a path toward fault-tolerant quantum computing by encoding logical qubits into larger systems of physical qubits, enabling the detection and correction of errors during computation.
However, QEC typically requires substantial resources, posing a major challenge for its realization on physical hardware.

A promising technique to mitigate the resource overhead of QEC is post-selection, which strategically discards low-confidence computations, thereby achieving significantly higher reliability from the remaining accepted results.
This approach enables attaining very low logical error rates with relatively small codes, with the trade-off of non-determinism.

Post-selection can be particularly useful for offline resource state generation processes such as magic state preparation \cite{bravyi2005universal,bravyi2012magic}, where aborting and retrying failed attempts does not damage existing encoded quantum information.
Additionally, when estimating expectation values of observables (e.g., ground-state energies via the variational quantum eigensolver \cite{peruzzo2014variational,bharti2022noisy}), discarding low-confidence runs can substantially improve estimator fidelity at the cost of additional sampling in the context of quantum error mitigation \cite{bonetmonroig2018low,sagastizabal2019experimental,cai2021quantum,tsubouchi2023virtual,lee2023error,cai2023quantum,zhou2025error,dinca2025error}.

In terms of QEC, initial approaches to post-selection were simple: abort the computation whenever any error was detected through syndrome measurements \cite{knill2005quantum,aliferis2008accuracy}.
This error-detection-based strategy could substantially improve effective error thresholds, but suffers from significant retry overhead costs.

Motivated by these initial works, QEC researchers have explored more sophisticated ``partial'' post-selection strategies that abort conditionally based on specific criteria even when errors are detected.
Namely, decoders produce \textit{soft outputs} (such as likelihoods) besides final corrections, which may contain information on the reliability of the decoded outcomes and thus can be employed to establish such criteria.
For example, an experimental work \cite{chen2022calibrated} demonstrated a reduction in the logical error rate by aborting when a correction from a decoder contains ambiguous faults.

The \textit{logical gap} (defined as the log-likelihood ratio difference between candidate corrections in distinct logical classes) has emerged as a Bayesian-motivated metric quantifying decoding confidence \cite{bombin2024fault,gidney2025yoked,smith2024mitigating,chen2025scalable}.
Post-selection strategies based on the logical gap proved remarkably effective, enabling about 15-fold reductions in error rates with a relative overhead factor of $<2$ \cite{bombin2024fault}, error threshold improvements (up to 50\% with full post-selection under code capacity noise) \cite{smith2024mitigating}, and order-of-magnitude reductions in magic state preparation costs \cite{gidney2024magic,lee2025low}.
The logical gap has become the standard approach for confidence-based post-selection in QEC.

Nevertheless, despite these advances, the logical gap method faces fundamental limitations that restrict broader applicability.\label{para:logical_gap_limitations}
First, computational overhead scales exponentially with the number $k$ of logical qubits (requiring comparative decoding over up to $4^k$ logical classes), which makes the method prohibitive for codes or circuits involving multiple logical qubits.
Second, the method is not guaranteed to work well for general decoders beyond the optimal degenerate maximum likelihood decoder \cite{iyer2015hardness,fuentes2021degeneracy} or the minimum-weight perfect matching (MWPM) decoder \cite{edmonds1965paths,higgott2022pymatching,higgott2025sparse}, as a correction within a fixed logical class may fail to represent the likelihood of the class.

Recent efforts have sought to address some of these limitations through alternative strategies, including decoder-independent confidence metrics based on syndrome density \cite{english2025thresholds} and cluster-geometry-based soft outputs \cite{meister2024efficient}.
However, these approaches still face challenges: the syndrome density method, while efficient and generalizable across code families, has sub-optimal performance, and the cluster-geometry-based method, while overcoming the exponential overhead problem of the logical gap, is currently applicable only to the repetition and surface codes.

In this work, we develop computationally efficient heuristic confidence metrics that overcome the limitations of the logical gap method, while maintaining effective post-selection performance across diverse quantum low-density parity check (QLDPC) codes.
Specifically, we quantify decoding confidence based on the clustering structure of errors revealed by clustering-based decoders, such as the union-find decoder \cite{delfosse2021almost,delfosse2022toward}, the belief propagation plus localized statistics decoder (BP+LSD) \cite{hillmann2025localized}, and the ambiguity clustering decoder \cite{wolanski2025ambiguity}.
Leveraging the intuition that the size and distribution of error clusters directly correlate with difficulty in decoding, we introduce two families of heuristic confidence metrics: the cluster size norm fraction and the cluster log-likelihood ratio (LLR) norm fraction.
Unlike the logical gap, our approach requires only a single decoder execution and is applicable to general QLDPC codes.

We demonstrate the effectiveness of our cluster-based confidence metrics through extensive numerical simulations on surface codes \cite{bravyi1998quantum,dennis2002topological}, bivariate bicycle codes \cite{kovalev2013quantum, bravyi2024high}, and hypergraph product codes \cite{improved2012kovalev,tillich2014quantum,zeng2019higher} using the BP+LSD decoder.
The results clearly show orders of magnitude reduction in logical error rates at moderate abort rates for all the three code families.
While our strategy does not match the logical gap method for surface codes (unsurprising given its strong probability-theoretical foundation), our approach's efficiency and generalizability make it a compelling alternative in practice.

Furthermore, we develop a real-time post-selection strategy based on the sliding-window decoding framework \cite{dennis2002topological, skoric2023parallel, tan2023scalable, huang2024increasing, gong2024toward,berent2024analog,scruby2024highthreshold,hillmann2025localized, kang2025quits}, which enables mid-circuit abort decisions that eliminate unnecessary overheads.
This strategy exhibits performance that is comparable to (or even better than) the global strategy in terms of the trade-off relation between the logical error rate and the average time cost.
Moreover, unlike the global strategy, our real-time strategy maintains a consistent per-round logical error rate and abort rate, which is a favorable property for circuits with large depths.

The main contributions of this work are:
\begin{itemize}
    \item We introduce cluster-based confidence metrics that overcome the computational and applicability limitations of the logical gap method, enabling efficient post-selection for general QLDPC codes.
    \item We develop a real-time post-selection strategy that integrates naturally with sliding-window decoding, allowing for early abort decisions and reduced computational overhead.
    \item We provide comprehensive numerical evidence demonstrating the effectiveness of our approach across multiple code families, achieving substantial reductions in logical error rates with modest abort rates.
\end{itemize}

\section{Results}

\subsection{Soft outputs from decoding}

The minimum requirement of a decoder is to infer which logical correction should be applied to counteract errors, given the detector outcomes.
Here, detectors indicate specific products of measurement outcomes deterministic in the absence of errors, which are normally constructed from two consecutive measurements of a stabilizer.
A detector error model (DEM) \cite{gidney2021stim} specifies the prior probabilities of independent fault locations (referred to as ``error mechanisms'') and the detectors flipped by each of them.
Decoders primarily aim to address the classification problem: given a DEM and detector outcomes, which logical class the current state belongs to.
In practice, however, decoders typically produce additional information beyond this basic classification.
These additional outputs are referred to as the \textit{soft outputs} of the decoder.

One common and straightforward soft output is the \emph{log-likelihood weight} of the correction, defined as the summation of the prior log-likelihood ratios (LLRs) for individual error mechanisms in the correction obtained from the decoder.
This approach is possible for many decoders, which return not only the logical class but also an explicit error correction (candidate error configuration) within the class.
Specifically, if $p_e$ is the prior probability of each error mechanism $e$, the log-likelihood weight of a correction $\tilde{E}$ is given by
\begin{align}
    w(\tilde{E}) \coloneqq \sum_{e \in \tilde{E}} \log\frac{1-p_e}{p_e}. \label{eq:correction_weight}
\end{align}
In particular, the MWPM decoder \cite{edmonds1965paths,higgott2022pymatching,higgott2025sparse} identifies a correction that minimizes this log-likelihood weight, which corresponds to a maximum likelihood solution when degeneracy is not considered.

Other useful soft outputs include \textit{posterior LLRs} of individual error mechanisms for given detector outcomes, which can be estimated using the belief propagation (BP) decoder \cite{mackay1996near}.
Due to inherent degeneracy of QEC codes, BP sometimes fails to converge and therefore does not exhibit an error threshold \cite{panteleev2021degenerate,raveendran2021trapping}.
Nevertheless, the posterior LLRs can be used in subsequent post-processing routines such as the ordered statistics decoder (OSD) \cite{roffe2020decoding, panteleev2021degenerate} or for running other decoders such as MWPM with updated weight information \cite{higgott2023improved}.

Here we focus on a specific application of soft outputs: quantifying decoding confidence.
Specifically, we aim to estimate the reliability of the logical flip inferred by a decoder through analysis of its soft outputs.
Given such a confidence metric, we construct a post-selection strategy that accepts a trial only when the metric value exceeds a specified cutoff threshold.
We note that soft outputs have found broader applications beyond post-selection, including hierarchical code architectures where confidence information from inner codes guides the decoding of outer codes \cite{gidney2025yoked, meister2024efficient,srivastava2025sequential}, construction of fast and accurate decoding framework by combining weak and strong decoders \cite{toshio2025decoder}, and error mitigation for logical circuits \cite{zhou2025error,dinca2025error,aharonov2025syndrome}.
However, this work focuses specifically on their application to post-selection.

\subsubsection{Logical gap for quantifying decoding confidence}
\label{sec:logical_gap}

A prominent approach in the literature for quantifying decoding confidence involves \emph{comparative} decoding and evaluation of the \textit{logical gap} (also known as the \textit{complementary gap}) \cite{bombin2024fault,gidney2025yoked,smith2024mitigating}.
Here, comparative decoding refers to a technique that performs decoding within each logical class and selects the minimum-weight correction among all classes.
The logical gap is then defined as the difference between the two smallest log-likelihood weights of corrections in distinct logical classes.

\label{para:logical_class_notion}
To formalize the notion of a logical class, we fix an ordered set $\mathcal{L} = \{\bar{Q}_1,\ldots,\bar{Q}_l\}$ of $l$ nontrivial independent logical Pauli errors.
Each class is then labeled by a \emph{logical flip vector} $\va{\lambda} \in \mathbb{Z}_2^l$, representing the logical effect of errors belonging to the class:
\begin{align*}
    \bar{P}\qty(\va{\lambda}) \coloneqq \prod_{j=1}^l \qty(\bar{Q}_j)^{\lambda_j}.
\end{align*}
Typically, $\mathcal{L}$ includes logical $X$ and $Z$ operators ($\lgx$, $\lgz$) for each of $k$ logical qubits; however, some may act trivially on the circuit (e.g., a $\lgz$ error just before a $\lgz$ measurement) or may not be independent (e.g., $\lgz_1$ and $\lgz_2$ just before a $\lgz_1\lgz_2$ measurement), so $l \le 2k$ in general.
In our numerical analysis, we consider circuits involving only $\lgz$ initialization/measurement, yielding $l = k$.

Comparative decoding aims to determine a correction \emph{conditioned} on a chosen logical class $\va{\lambda}$.
Operationally, this means we want to restrict decoding to errors whose net logical effect equals $\bar{P}(\va{\lambda})$.
We implement this conditioning by augmenting the DEM with $l$ additional \emph{virtual detectors}, one for each $\bar{Q}_j \in \mathcal{L}$, and then fixing their values to $\lambda_j$ during decoding.
Intuitively, a virtual detector is an auxiliary syndrome bit that records whether a conjugate logical observable $\bar{R}_j$ would be flipped; by pinning this bit to $\lambda_j$, we force the decoder to output a correction in the desired logical class.

Formally, for each $\bar{Q}_j \in \mathcal{L}$ we choose a logical Pauli operator $\bar{R}_j$ anticommuting with $\bar{Q}_j$ and commuting with all other elements in $\mathcal{L}$.
We then (conceptually) measure $\bar{R}_j$ immediately after the decoded circuit, obtaining an outcome from the product of certain physical measurement outcomes.
The virtual detector associated with $\bar{Q}_j$ is defined to flip by every error mechanism that flips this product.\footnote{
    If a logical measurement of $\bar{R}_j$ is already present in the circuit, we use it directly; otherwise, we temporarily append an \emph{ideal} circuit fragment that measures $\bar{R}_j$ purely for the purpose of constructing the augmented DEM.
    Practically, this can be automated in \emph{Stim} using \texttt{OBSERVABLE\_INCLUDE}: (1) Wrap the circuit with the required logical $\bar{R}_j$ initialization/measurement layers when absent, (2) mark the corresponding measurement-outcome product as an observable via \texttt{OBSERVABLE\_INCLUDE}, (3) generate the DEM, and (4) relabel the resulting observable as an additional detector so that it can be fixed to $\lambda_j$ during decoding.
    The temporary $\bar{R}_j$ measurement is an auxiliary construction and need not be included in actual comparative decoding runs.
}
Decoding the augmented DEM with the virtual detectors fixed to $\va{\lambda}$ therefore yields a correction \emph{within} the logical class $\va{\lambda}$.
Note that virtual detectors generalize the notion of boundary detectors in the surface-code setting \cite{gidney2025yoked,meister2024efficient}.

After performing comparative decoding over all logical classes, the logical gap $\Delta$ is defined as the difference between the two smallest log-likelihood weights among the corrections $\{ \tilde{E}_{\va{\lambda}}\}_{\va{\lambda} \in \mathbb{Z}_2^{l}}$ in different logical classes, i.e., 
\begin{align} 
    \Delta \coloneqq w_{(2)} - w_{(1)}, \label{eq:logical_gap_def}
\end{align} 
where $w_{(1)} \le w_{(2)} \le \cdots \le w_{(2^{l})}$ are the sorted values of $[w(\tilde{E}_{\va{\lambda}})]_{\va{\lambda} \in \mathbb{Z}_2^{l}}$. 
A large logical gap indicates that error patterns in alternative logical classes are significantly less likely than the predicted one, thereby implying high confidence in the decoding outcome.

\label{para:extending_logical_gap_difficulty}
Despite this general description, the logical gap method has been investigated almost exclusively for the minimum-weight perfect matching (MWPM) decoder.
To understand why extending it to other decoders is fundamentally difficult, we first outline the core requirements that make the logical gap meaningful.

The logical gap quantifies how much more likely the predicted logical class is compared to alternatives.
For this comparison to be valid, two conditions must hold:
(i) The decoder must find a correction that faithfully represents the likelihood of each logical class, and
(ii) the decoder must maintain its normal performance when constrained to decode within a specific logical class.
The MWPM decoder satisfies both conditions: It can identify the minimum-weight (i.e., maximum-likelihood under independent noise) correction within any logical class, and its matching algorithm operates identically regardless of which logical class is specified.
Other efficient decoders generally fail one or both of these requirements, as we elaborate below.\footnote{
    We here focus on decoders with polynomial-time complexity, which are essential for practical QEC.
    In principle, computationally expensive decoders such as the degenerate quantum maximum likelihood decoder \cite{iolius2024decoding} can assess decoding confidence even more reliably than MWPM by evaluating the true likelihood of each logical class rather than approximating it via a single representative correction.
    However, their exponential computational cost in the general case renders them impractical for real-time decoding, despite recent progress on polynomial-time algorithms for restricted settings \cite{cau2025exact}.
}

First, most efficient decoders besides MWPM do not guarantee finding the minimum-weight correction within a fixed logical class.
When the decoder returns a suboptimal correction for a given class, the associated weight no longer accurately reflects that class's likelihood.
Consequently, comparing weights across classes becomes unreliable: A logical class might appear less likely simply because the decoder failed to find a good correction for it, not because errors in that class are genuinely improbable.
This undermines the fundamental premise of the logical gap as a confidence measure.

Second, constraining a decoder to a specific logical class requires introducing ``virtual'' detectors that enforce the logical constraint (as described above).
These virtual detectors can severely disrupt the internal heuristics of many decoders.
For instance, the union-find (UF) decoder \cite{delfosse2021almost} grows clusters incrementally from violated detectors.
A virtual detector, however, connects to $\Omega(d)$ error mechanisms (where $d$ is the code distance), causing any cluster touching it to grow abnormally fast and merge with distant clusters.
This destroys the locality structure that the UF decoder exploits, leading to degraded decoding quality.
The same issue affects generalized UF decoders for quantum low-density parity-check (QLDPC) codes \cite{delfosse2022toward}, which rely on low-degree connectivity that high-degree virtual detectors violate.

Due to these limitations, the logical gap method is generally not compatible beyond the MWPM decoder.
However, the MWPM decoder is applicable only to a restricted class of QEC codes; specifically, those where each elementary error mechanism affects at most two detectors, such as repetition and surface codes \cite{iolius2024decoding}.
A notable exception is the class of two-dimensional color codes \cite{bombin2006topological}, which are not directly compatible with MWPM but admit suboptimal decoders that apply MWPM to specific matchable subgraphs \cite{delfosse2014decoding,chamberland2020triangular,beverland2021cost,sahay2022decoder,kubica2023efficient,gidney2023new,lee2025color}.
For these decoders, the logical gap method remains reasonably effective \cite{lee2025low}, particularly with the concatenated MWPM decoder \cite{lee2025color}.

Last but not least, the logical gap method may be impractical due to its computational cost, which scales exponentially with the number $l$ of nontrivial independent logical Pauli errors (at most $2k$ for $k$ logical qubits), as it requires decoding across all $2^{l}$ logical classes.
While manageable for memory experiments with a single logical qubit, this scaling renders the method difficult to use for multiple logical qubits with correlated noise.
This limitation can be particularly fatal for QLDPC codes, which typically encode multiple logical qubits per code block; for instance, a single code block of the $[[144, 12, 12]]$ bivariate bicycle code \cite{kovalev2013quantum, bravyi2024high} has at most $4^{12} = \num{16777216}$ logical classes.

\subsection{Post-selection strategy based on error cluster statistics}

Due to the aforementioned limitations of the logical gap method, we seek alternative approaches for quantifying decoding confidence and constructing post-selection strategies.
This requires decoders that provide rich soft outputs about error configurations, a condition that clustering-based decoders satisfy well.
These decoders operate by inferring the clustering structure of errors (the pattern of how errors group together) which could directly correlate with the difficulty of the decoding process.

\subsubsection{Clustering-based decoders}
\label{sec:clustering_based_decoders}

To briefly introduce relevant terms first, the \textit{check matrix} is a binary matrix indicating how error mechanisms (columns) and detectors (rows) are connected, which forms the biadjacency matrix of the bipartite \textit{Tanner graph} with check nodes (corresponding to detectors) and fault nodes (corresponding to error mechanisms).
A \textit{cluster} is a set of error mechanisms that forms a connected region in the \textit{fault graph} composed of only fault nodes, where two fault nodes are connected if the corresponding error mechanisms are commonly involved in at least one detector.
A cluster is considered \textit{valid} if it contains at least one solution consistent with all detectors that nontrivially involve any error mechanisms in the cluster.
Formal definitions of these graph-theoretic structures are provided in Methods.

\begin{figure*}[!t]
	\centering
	\includegraphics[width=\textwidth]{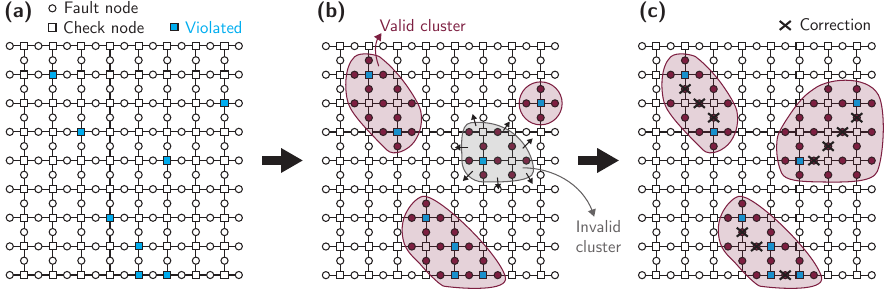}
	\caption{
        \textbf{General workflow of a clustering-based decoder.}
        \subfig{a} The tanner graph of the distance-9 surface code patch under bit-flip noise is shown with one possible example of detector outcome pattern.
        \subfig{b} A collection of disjoint valid clusters (which are guaranteed to have local solutions) are constructed according to a specific rule.
        The rule typically involves growing a cluster incrementally from each violated check node until it becomes valid, during which colliding clusters are merged.
        Note that while clusters are formally defined on the fault graph, we visualize them here on the Tanner graph since cluster growth rules naturally involve check node (detector) information.
        \subfig{c} Each cluster is decoded independently via a specific inner decoder.
        The local corrections are aggregated to form the final global correction.
        Note that the cluster construction and decoding processes shown here are for illustrative purposes only and do not correspond to any specific decoder.
    }
	\label{fig:clustering_based_decoder}
\end{figure*}

As illustrated in Fig.~\ref{fig:clustering_based_decoder}, clustering-based decoders generally operate through three main steps (see Definition~\ref{def:clustering_based_decoder} in Methods for a formal definition):
\begin{enumerate}
    \item \textit{Cluster construction}: The decoder constructs a collection of disjoint valid clusters according to a specific rule, such that the union of detectors associated with these clusters covers all violated detectors. Typically, a cluster is grown incrementally from each violated check node until it becomes valid, during which colliding clusters are merged.
    \item \textit{Local decoding}: Each cluster is decoded independently by solving a reduced decoding problem restricted to the error mechanisms within the cluster and the detectors they affect.
    \item \textit{Correction aggregation}: The local corrections obtained from each cluster are combined to form the global correction.
\end{enumerate}
The key differentiating factor among clustering-based decoders lies in how they organize the three components: (i) cluster growth, (ii) validity checking, and (iii) local decoding within each cluster.

The union-find (UF) decoder \cite{delfosse2021almost} is a representative example of this approach applicable to topological codes.
In terms of the three components: (i) clusters are grown and fused incrementally from violated detectors using a union-find data structure, (ii) validity is checked by simple parity counting, and (iii) local decoding is performed via the peeling decoder which identifies a spanning tree for each cluster.
The UF decoder was later generalized to quantum low-density parity-check (QLDPC) codes \cite{delfosse2022toward}.

\label{para:clustering_based_decoders_another_class}
Another class of clustering-based decoders leverages belief propagation (BP) outcomes to guide cluster formation, prioritizing high-likelihood error mechanisms; examples include BP plus localized statistics decoding (BP+LSD) \cite{hillmann2025localized} and ambiguity clustering (AC) \cite{wolanski2025ambiguity}.
Specifically, the BP+LSD decoder first runs BP, accepting its solution if it converges, and otherwise applying LSD postprocessing.
In LSD, (i) clusters are grown incrementally by including adjacent error mechanisms with highest BP likelihoods and (ii), (iii) a matrix factorization algorithm called on-the-fly elimination enables efficient validity check and local decoding \cite{roffe2020decoding,panteleev2021degenerate}.
We provide a detailed description of BP+LSD in Methods, as it serves as the primary decoder throughout our numerical analysis.

\subsubsection{Heuristic confidence metrics from cluster statistics}

\begin{figure*}[!t]
	\centering
	\includegraphics[width=\textwidth]{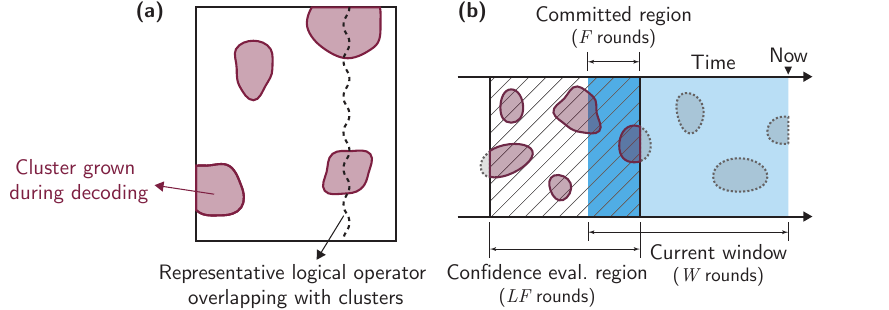}
	\caption{
        \textbf{Overview of our heuristic cluster-based confidence metrics and post-selection strategies based on them.}
        \subfig{a} After constructing valid clusters from a clustering-based decoder, cluster statistics are used to quantify decoding confidence.
        As the size and number of clusters grow, a larger portion of the logical operator’s support tends to overlap with clusters, thereby increasing the probability of logical failure.
        Based on this intuition, we define two metrics (the cluster size and LLR norm fractions) in Definitions~\ref{def:cluster_size_norm_fraction} and~\ref{def:cluster_llr_norm_fraction}, and introduce post-selection strategies that abort a trial when the metric value exceeds a chosen cutoff.
        \subfig{b} This approach is further extended to real-time decoding with the sliding-window framework.
        After decoding each window of $W$ rounds and committing its first $F$ rounds, clusters are constructed from committed errors within the latest $LF$ rounds (where $L$ denotes the lookback window size), and a metric is evaluated based on them.
        If the value exceeds the cutoff, the trial is immediately aborted, thereby avoiding unnecessary computation.
    }
	\label{fig:overview}
\end{figure*}

We construct heuristic confidence metrics by leveraging cluster statistics obtained from clustering-based decoders, such as cluster sizes and aggregated error probabilities within clusters.
Intuitively, smaller error clusters indicate better-localized error configurations in terms of the connectivity structure on the Tanner graph, resulting in less ambiguity during error inference.
As the size and number of clusters increase, the portion of a logical operator’s support that lies outside the clusters will shrink, and therefore the probability of decoding failure (i.e., the probability that a logical error occurs due to an error chain existing outside the clusters) will increase; see Fig.~\ref{fig:overview}(a).

Based on these intuitions, we propose two families of heuristic confidence metrics (strictly, ``inverse'' confidence metrics), parametrized by a positive number $\alpha$ (that can be infinity $\infty$): the \textit{cluster size $\alpha$-norm fraction} and the \textit{cluster LLR $\alpha$-norm fraction}, defined as follows:

\begin{definition}[Cluster size $\alpha$-norm fraction]
    For a set of error mechanisms $\mathcal{E}$, disjoint clusters $\qty{C_i}_i$ (such that $\forall i, C_i \subseteq \mathcal{E}$), and a positive real number $\alpha$, the cluster size $\alpha$-norm fraction is defined as
    \begin{align*}
        \sizeNormFrac{\alpha}\qty(\qty{C_i}_i; \mathcal{E}) := \frac{1}{\abs{\mathcal{E}}}\qty(\sum_i \abs{C_i}^\alpha)^{1/\alpha} \in [0, 1],
    \end{align*}
    where $\abs{\mathcal{E}}$ denotes the number of elements in $\mathcal{E}$.
    The $\alpha=\infty$ case is additionally defined as
    \begin{align*}
        \sizeNormFrac{\infty}\qty(\qty{C_i}_i; \mathcal{E}) := \frac{1}{\abs{\mathcal{E}}}\max_i \abs{C_i} \in [0, 1].
    \end{align*}
    \label{def:cluster_size_norm_fraction}
\end{definition}

\begin{definition}[Cluster LLR $\alpha$-norm fraction]
    For a set of error mechanisms $\mathcal{E}$, disjoint clusters $\qty{C_i}_i$ (such that $\forall i, C_i \subseteq \mathcal{E}$), and a positive real number $\alpha$, the cluster LLR $\alpha$-norm fraction is defined as
    \begin{align*}
        \llrNormFrac{\alpha}\qty(\qty{C_i}_i; \mathcal{E}) := \frac{1}{\sum_{e \in \mathcal{E}} w_e}\qty[\sum_i \qty(\sum_{e \in C_i} w_e)^\alpha]^{1/\alpha} \in [0, 1],
    \end{align*}
    where $w_e := \log[(1-p_e)/p_e]$ is the prior LLR of error mechanism $e$.
    The $\alpha=\infty$ case is additionally defined as
    \begin{align*}
        \llrNormFrac{\infty}\qty(\qty{C_i}_i; \mathcal{E}) := \frac{1}{\sum_{e \in \mathcal{E}} w_e} \max_i \sum_{e \in C_i} w_e \in [0, 1].
    \end{align*}
    \label{def:cluster_llr_norm_fraction}
\end{definition}

These metrics are expected to correlate inversely with decoding confidence.
The cluster LLR $\alpha$-norm fraction ($\llrNormFrac{\alpha}$) additionally weights errors by their LLRs, reflecting that clusters with lower LLR values (higher error probabilities) are more likely to contain real errors.
The parameter $\alpha$ controls the influence of large clusters: As $\alpha$ increases, larger clusters dominate the metric, with the limiting case $\alpha=\infty$ considering only the largest cluster.

The set of error mechanisms $\mathcal{E}$ in the definitions may be the full error mechanism set of a DEM, or a subset when disjoint groups of mechanisms are decoded independently. 
For instance, in Calderbank–Shor–Steane (CSS) codes, if circuits involve only logical $Z$ resets and measurements, it suffices to take $\mathcal{E}$ as the set of $X$ errors to capture confidence information, provided that correlations between $X$ and $Z$ errors are ignored.
This approach prevents errors unrelated to the logical Pauli correction under decoding from affecting its confidence evaluation.

Note that previous works \cite{meister2024efficient,smith2024mitigating} have proposed cluster-based confidence metrics as well, based on the weight of the shortest segment among all nontrivial error strings that does not overlap with any cluster.
Our approach differs by relying solely on information intrinsic to each cluster, without requiring explicit knowledge of the supports of logical operators (whose exhaustive search can be computationally expensive for general codes).

\subsubsection{Global post-selection strategy}

Based on the cluster norm fraction metrics in Definitions~\ref{def:cluster_size_norm_fraction} and~\ref{def:cluster_llr_norm_fraction}, we propose the following post-selection strategy (the term ``global decoding'' is used to distinguish from the real-time strategy):

\begin{strategy}[Cluster-based post-selection for global decoding]
    Execute a clustering-based decoder and collect cluster size or LLR data (optionally restricted to a subset of error mechanisms that nontrivially affects the logical Pauli correction under decoding). Given a method $m \in \qty{\mathrm{size}, \mathrm{LLR}}$, norm order $\alpha > 0$, and cutoff threshold $c \in [0, 1]$, accept the decoding result only when $Q_m^{(\alpha)} \leq c$.
    \label{strategy:global}
\end{strategy}

This strategy offers several advantages over the logical gap method:
\begin{enumerate}
    \item Some clustering-based decoders such as the generalized UF, BP+LSD, and AC decoders are applicable to general QLDPC codes.
    \item No comparative decoding is required; a single decoder execution suffices to compute the metrics, while computing the logical gap requires at most $4^{k}$ executions for $k$ logical qubits.
    \item The approach integrates naturally with real-time modular decoding (e.g., sliding-window methods \cite{dennis2002topological, skoric2023parallel, tan2023scalable, huang2024increasing, gong2024toward,berent2024analog,scruby2024highthreshold,hillmann2025localized, kang2025quits}), as global cluster statistics can be inferred from partial information in intermediate decoding outcomes.
    \item Abort decisions can be made after constructing valid clusters but before running local decoders, avoiding unnecessary computational overheads when aborting. However, this advantage may be modest since cluster construction is often more computationally heavy than its decoding. For instance, the BP+LSD decoder requires $O(n^3)$ basic operations to build a valid cluster while $O(n^2)$ to decode it, where $n$ is the cluster size \cite{hillmann2025localized}.
\end{enumerate}

\subsection{Real-time post-selection strategy}

While global decoders that process all syndrome data at once are useful for proof-of-principle QEC analysis, they can be impractical for real quantum computing systems.
In practice, feedforward corrections must be computed in real time before executing any non-Clifford gate, as such corrections may involve Clifford operations that cannot be absorbed into classical Pauli frame updates.
This requirement necessiates real-time decoding, which needs to be sufficiently fast to prevent the backlog problem that could lead to an exponential slowdown during computation \cite{terhal2015quantum,skoric2023parallel,barber2025real,caune2024demonstrating}.

We extend our post-selection strategy to a real-time setting by integrating it with the sliding-window decoding framework for general QLDPC codes \cite{huang2024increasing, gong2024toward,berent2024analog,scruby2024highthreshold,hillmann2025localized, kang2025quits}.
Rather than collecting clusters statistics from global decoding outcomes, we rely only on data from a constant number of recent windows as shown in Fig.~\ref{fig:overview}(b), enabling us to compute a cluster-confidence metric in real time and to triger early aborts when necessary.

The sliding window decoding framework decomposes the overall decoding problem into smaller, overlapping sub-problems defined on ``windows'', which can be solved incrementally.
By partitioning along the time direction, a partial correction can be derived from the syndrome data collected so far, enabling real-time decoding.
The algorithm is characterized by two positive integers $(W, F)$: the window size $W$ (which governs decoding speed and needs to be small enough to avoid the backlog problem) and the commit size $F$ (which corresponds to the width of overlapping regions between adjacent windows).
See the Methods section for more details on the algorithm.

To integrate our post-selection strategy with the sliding-window framework, we employ a clustering-based decoder as the inner decoder.
As described in Fig.~\ref{fig:overview}(b), we track fractions of clusters overlapping with committed regions and evaluate a cluster norm fraction metric based on the $L$ most recent committed regions.
Importantly, whether to abort is judged after decoding each window, potentially reducing the retry cost.
The detailed strategy is as follows:
\begin{strategy}[Cluster-based post-selection for real-time sliding-window decoding]
    Let $L \geq 1$ be the lookback window size, $m \in \{\mathrm{size}, \mathrm{LLR}\}$ the cluster weighting method, $\alpha > 0$ the norm order, and $c \in [0, 1]$ the cutoff threshold.
    After decoding each window $w \geq L - 1$:
    \begin{enumerate}
        \item Record the clusters $\qty{C_{w,i}}_i$ obtained from the current window.
        \item Identify all committed error mechanisms that lie within clusters from the last $L$ windows:
        \begin{align*}
            \mathcal{E}_{L,w}^\mathrm{inside} \coloneqq \bigcup_{w'=w-L+1}^w \qty(\bigcup_i C_{w',i} \cap \mathcal{E}_{w'}^{\text{commit}}).
        \end{align*}
        Here, $\mathcal{E}_{w}^{\text{commit}}$ is defined as the set of error mechanisms committed from window $w$; see Eq.~\eqref{eq:committed_error_mechanisms} in the Methods section for the formal definition.
        \item Construct ``committed clusters'' $\{C_{L,w,i}^\mathrm{commit}\}_i$ by grouping the error mechanisms in $\mathcal{E}_{L,w}^\mathrm{inside}$ into connected components within the fault graph.
        \item Compute the cluster norm fraction 
        $$Q \coloneqq Q_m^{(\alpha)}\qty(\qty{C_{L,w,i}^\mathrm{commit}}_i; \bigcup_{w'=w-L+1}^w \mathcal{E}_{w'}^{\text{commit}})$$ 
        for these committed clusters.
        \item If $Q > c$, abort the trial immediately.
        \item Otherwise, proceed to decode the next window and repeat this process until either it is aborted or the circuit ends.
    \end{enumerate}
    \label{strategy:realtime}
\end{strategy}

\subsection{Performance analysis}

We present a comprehensive numerical evaluation of the post-selection strategies introduced in Strategies~\ref{strategy:global} and~\ref{strategy:realtime}.
Our simulations employ the BP+LSD-0 decoder \cite{hillmann2025localized} for collecting cluster statistics, which is chosen for its broad applicability to arbitrary QLDPC codes, its error-correcting performance comparable to the BP+OSD decoder \cite{roffe2020decoding,panteleev2021degenerate}, and its availability through the open-source Python package \texttt{ldpc} \cite{roffee2022ldpc}.
To ensure cluster statistics are available for all samples, we modified the decoder to execute the LSD subroutine even when BP converges successfully.
All simulations use 30 BP iterations with the min-sum method; further optimization of these parameters could potentially enhance the numerical results.
Details on our simulation methods are presented in the Methods section.

\subsubsection{Global strategy analysis \label{subsubsec:global_strategy_analysis}}

We conduct numerical simulations using memory experiments with $T=d$ rounds of syndrome extraction for logical $Z$ observables. 
Our analysis covers three representative code families: rotated surface codes, bivariate bicycle (BB) codes \cite{kovalev2013quantum, bravyi2024high}, and hypergraph product (HGP) codes \cite{improved2012kovalev,tillich2014quantum,zeng2019higher}, with code distance $d$.

The circuits are subjected to the uniform circuit-level noise model with error strength $p$, where each single-qubit gate, two-qubit gate, or idling is followed by depolarizing errors with probability $p$, and resets/measurements are flipped with the same probability.
After decoding, we compute cluster size and LLR norm fractions with norm orders $\alpha \in \qty{0.5, 1, 2, \infty}$, restricted to error mechanisms associated with $Z$-type detectors.
Strategy~\ref{strategy:global} is used for determining acceptance, and the logical error rate $\plog$ (at which any logical observable is erroneous) is evaluated over accepted samples.

Our primary focus is quantifying the trade-off between $\plog$ and the abort rate $\pabort$: \textit{How much can the logical error rate be reduced by aborting a certain fraction of low-confidence samples?}\label{para:baseline_metrics}
Since no established confidence metric exists for general QLDPC codes beyond the logical gap via comparative decoding, we compare against two baseline metrics that represent fundamental soft outputs applicable to any decoder with minimal computational overhead:
\begin{itemize}
    \item \textbf{Correction weight}: The log-likelihood weight of the decoder's correction, as defined in Eq.~\eqref{eq:correction_weight}.
    This quantity is the cost function that (non-degenerate) maximum likelihood decoders including MWPM directly minimize, and is related to error likelihood via $\Pr(E) \propto \exp(-w(E))$.
    High correction weight indicates an atypical, noisy shot, suggesting lower confidence in the decoded result.
    \item \textbf{Detector density}: The fraction of violated detectors relative to the total detector count.
    Like correction weight, detector density quantifies how noisy and atypical a shot is.
    Unlike other metrics, however, it can be computed before decoder execution, offering exceptional computational efficiency (though it is not strictly a decoder soft output).
    This metric is related to the annular-syndrome post-selection rule in Ref.~\cite{bombin2024fault} and extends the ``non-equilibrium magnetization'' heuristic from Ref.~\cite{english2025thresholds} (originally developed for code-capacity noise) to circuit-level noise, motivated by the insight that syndrome density serves as an order parameter in the associated statistical-mechanical spin model~\cite{english2025thresholds, chubb2021statistical}.
\end{itemize}
Both heuristic baselines are expected to have inverse correlation with decoder confidence, which we confirm numerically.

Additionally, we compare with the logical gap, which requires comparative decoding across logical classes. \label{para:logical_gap_analysis}
For surface codes (with a single logical qubit), this is computationally tractable, thus we evaluate the logical gap using both MWPM and BP+LSD to assess whether the method extends beyond MWPM decoders.
For BB and HGP codes where exhaustive comparative decoding requires enormous computational resources, we try approximating the logical gap (referred to as the ``logical gap proxy'') by decoding within a subset of $N_\mathrm{class}$ classes composed of the standard BP+LSD correction and $N_\mathrm{class} - 1$ other logical classes selected randomly.
Specifically, we use $N_\mathrm{class} = 24$ for the $[[144, 12, 12]]$ BB code and $N_\mathrm{class} = 18$ for the $[[225, 9, 6]]$ HGP code.
(We also test more sophisticated heuristics for the gap proxy based on the empirical logical error distribution, but none outperform uniform random sampling; see the Supplementary Material for details.)

\begin{figure*}[!t]
	\centering
	\includegraphics[width=\textwidth]{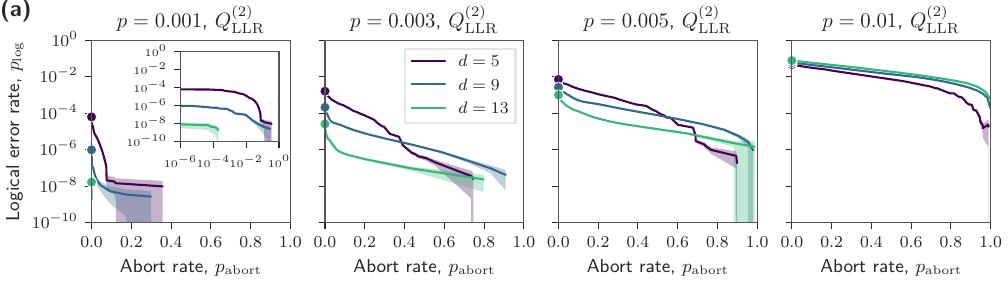}\\[0.5em]
    \includegraphics[width=0.32\textwidth]{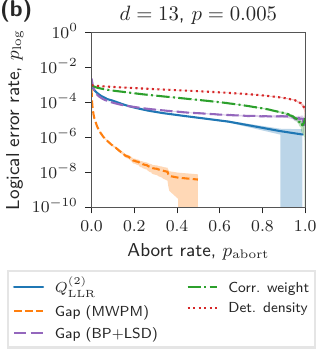}
    \includegraphics[width=0.32\textwidth]{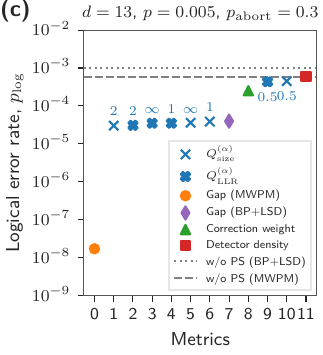}
    \includegraphics[width=0.32\textwidth]{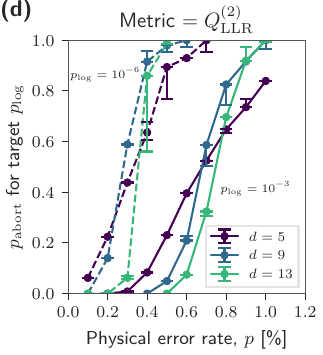}
	\caption{
        \textbf{Post-selection analysis of global decoding for rotated surface codes}.
        \subfig{a} Logical error rates $\plog$ are plotted against $\pabort$ for Strategy~\ref{strategy:global} based on the cluster LLR 2-norm fraction $\llrNormFrac{2}$, across different physical error rates $p \in \{0.001, 0.003, 0.005, 0.01\}$ and code distances $d \in \{5, 9, 13\}$. 
        The values of $\plog$ without post-selection ($\pabort = 0$) are emphasized as filled circles.
        For $p=0.001$, an inset displaying the same data with a logarithmic scale on $\pabort$ is included.
        Shaded regions indicate 95\% confidence intervals.
        \subfig{b} Our strategy based on $\llrNormFrac{2}$ is compared with four other strategies for $d=13$ and $p=0.005$: the logical gap calculated by MWPM, the logical gap calculated by BP+LSD, correction weight, and detector density.
        \subfig{c} Various strategies are compared at a fixed abort rate $\pabort = 0.3$.
        We consider four norm orders $\alpha \in \{0.5, 1, 2, \infty\}$ (specified above or below the markers) for the cluster size and LLR norm fractions.
        \subfig{d} Required abort rates $\pabort$ to achieve target values of $\plog \in \{10^{-3}, 10^{-6}\}$ are plotted against $p$ when using the metric $\llrNormFrac{2}$.
    }
	\label{fig:surface_code_simulation_results}
\end{figure*}

\paragraph*{Surface codes.}
Figure~\ref{fig:surface_code_simulation_results} presents comprehensive results for surface codes with distances $d \in \qty{5, 9, 13}$.
In Fig.~\ref{fig:surface_code_simulation_results}(a), we plot the trade-off between $\plog$ and $\pabort$ using the cluster LLR 2-norm fraction ($\llrNormFrac{2}$) across physical error rates $p \in \qty{0.001, 0.003, 0.005, 0.01}$. 
Note that the subplot for $p=0.001$ includes an inset displaying the same data with a logarithmic scale on $\pabort$, as $\plog$ varies rapidly near $\pabort=0$.
For most parameter regimes (particularly when $p \leq 0.005$), the strategy shows several orders of magnitude improvement in $\plog$ with modest abort rates.

\label{para:surface_code_results}
In Fig.~\ref{fig:surface_code_simulation_results}(b), we benchmark our cluster-based approach against established methods for $(d, p) = (13, 0.005)$, comparing $\llrNormFrac{2}$ with the logical gap computed via both MWPM and BP+LSD, correction weight, and detector density.
As expected, our method's performance falls between the near-optimal MWPM logical gap strategy and the simpler baseline metrics.
Notably, the BP+LSD logical gap exhibits significantly degraded performance compared to MWPM (achieving only $\plog \sim 10^{-5}$ versus $\sim 10^{-8}$ at $\pabort = 0.3$) despite using the same comparative decoding framework.
We attribute this degradation to BP+LSD not guaranteeing the minimum-weight correction within each logical class, which can cause the correction weights to misrepresent the true likelihood of each class (as discussed in the ``Logical gap for quantifying decoding confidence'' section).

Figure~\ref{fig:surface_code_simulation_results}(c) also compares various strategies, but at a fixed abort rate $\pabort=0.3$, with different values of the norm order $\alpha$ (specified above or below each data point) plotted separately for cluster-based metrics.
The results show that the choice of $\alpha$ has little impact on performance as long as $\alpha \geq 1$.
Lastly, Fig.~\ref{fig:surface_code_simulation_results}(d) examines the question of \textit{how much we need to abort to achieve a target $\plog$}, which may be more practically relevant.
The required abort rates to reach target values $\plog \in \qty{10^{-3}, 10^{-6}}$ via the strategy based on $\llrNormFrac{2}$ are plotted against $p$ across different code distances.

\begin{figure*}[!t]
	\centering
	\includegraphics[width=0.8\textwidth]{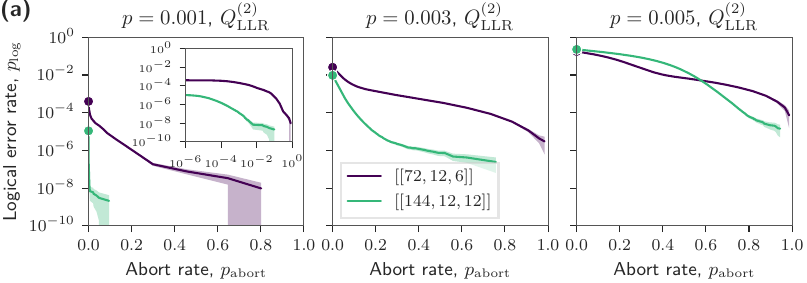}\\[0.5em]
    \includegraphics[width=0.32\textwidth]{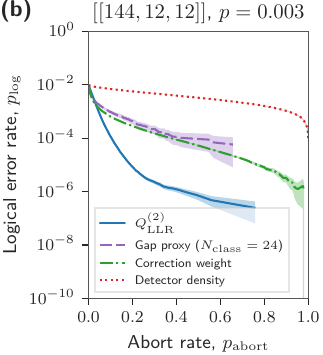}
    \includegraphics[width=0.32\textwidth]{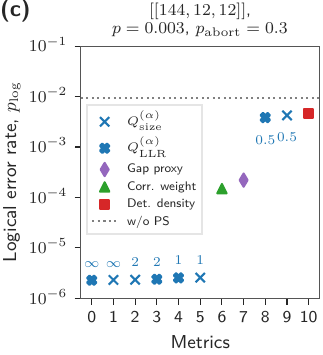}
    \includegraphics[width=0.32\textwidth]{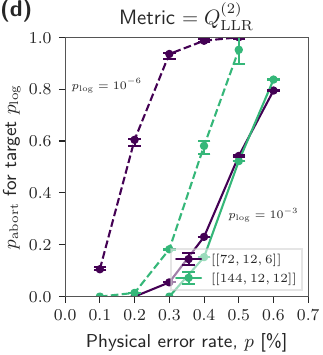}
	\caption{
        \textbf{Post-selection analysis of global decoding for bivariate bicycle codes.}
        Two variants of bivariate bicycle codes are considered: $[[144, 12, 12]]$ and $[[72, 12, 6]]$.
        Shaded regions represent 95\% confidence intervals.
        \subfig{a} Any-observable logical error rates $\plog$ are plotted against the abort rate $\pabort$ at $p \in \{0.001, 0.003, 0.005\}$ for Strategy~\ref{strategy:global} based on $\llrNormFrac{2}$.
        For $p = 0.001$, a log-scale inset is included.
        \subfig{b}, \subfig{c} Various strategies are compared for the $[[144, 12, 12]]$ BB code at $p = 0.003$, including a logical gap proxy (``random'' heuristic) calculated from a subset of logical classes composed of the standard BP+LSD correction and $N_{\text{class}} - 1 = 23$ other classes selected randomly.
        $\pabort$ is fixed to 0.3 in \subfig{c}.
        \subfig{d} Required $\pabort$ to achieve target values of $\plog \in \{10^{-3}, 10^{-6}\}$ are plotted against $p$.
    }
	\label{fig:bb_code_simulation_results}
\end{figure*}

\paragraph*{Bivariate bicycle codes.}
Figure~\ref{fig:bb_code_simulation_results} presents the simulation results for two instances of BB codes: $[[72, 12, 6]]$ and $[[144, 12, 12]]$ \cite{bravyi2024high}.
(Here, $[[n, k, d]]$ characterizes a code with $n$ physical qubits, $k$ logical qubits, and the code distance $d$.)
In Fig.~\ref{fig:bb_code_simulation_results}(a), $\plog$ is plotted against $\pabort$ at $p \in \qty{0.001, 0.003, 0.005}$, where the subplot for $p=0.001$ includes a log-scale inset.
Various strategies are compared in Fig.~\ref{fig:bb_code_simulation_results}(b) and~(c) for the $[[144, 12, 12]]$ code at $p=0.003$, confirming that the performance of the cluster-based strategy surpasses the correction weight strategy by up to around two orders of magnitude and does not severely depend on the norm order $\alpha$ as long as $\alpha \geq 1$.

\label{para:bb_code_logical_gap_proxy_results}
Additionally, the gap proxy method with $N_\mathrm{class}=24$ fails to outperform even the correction weight despite the (at least) 24-fold increase in computational cost compared to standard decoding.
This may be attributed to two factors: (i) The logical gap method itself is less effective for BP+LSD than for MWPM, as demonstrated by the surface code results above, and (ii) approximating the gap using only a subset of randomly sampled logical classes introduces additional estimation error, further degrading performance.

The required abort rates to reach target values $\plog \in \qty{10^{-3}, 10^{-6}}$ are plotted against $p$ in Fig.~\ref{fig:bb_code_simulation_results}(d).

\begin{figure*}[!t]
	\centering
    \includegraphics[width=0.35\textwidth]{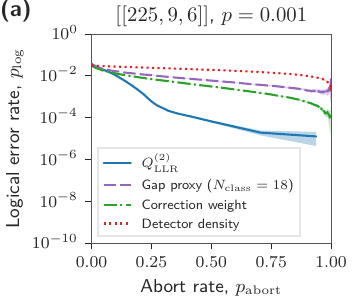}
    \includegraphics[width=0.64\textwidth]{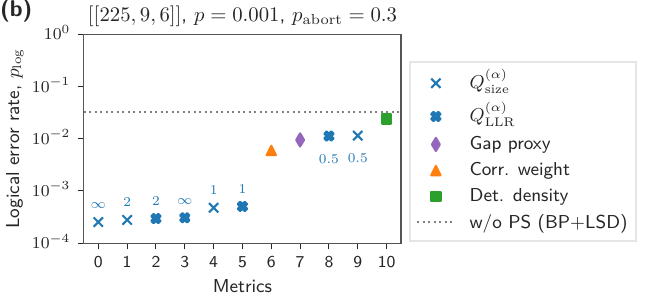}
	\caption{
        \textbf{Post-selection analysis of global decoding for a hypergraph product code.}
        We consider a $[[225, 9, 6]]$ $(3, 4)$-regular hypergraph product code, defined from the product of a 12-variable classical LDPC codes with itself.
        \subfig{a} Any-observable logical error rates $\plog$ are plotted against the abort rate $\pabort$ at $p = 0.001$, comparing Strategy~\ref{strategy:global} based on $\llrNormFrac{2}$ with the heuristic baselines and a gap proxy ($N_{\text{class}} = 18$, ``random'' heuristic).
        Shaded regions represent 95\% confidence intervals.
        \subfig{b} Various strategies are compared at a fixed $\pabort = 0.3$.
    }
	\label{fig:hgp_code_simulation_results}
\end{figure*}

\paragraph*{Hypergraph product codes.}
We evaluate our approach on a $(3, 4)$-regular HGP code with parameters $[[225, 9, 6]]$, constructed by taking the product of a 12-variable classical LDPC code with itself, whose check matrix is given in the Methods section.
This particular code offers practical advantages with its depth-8 syndrome extraction circuit \cite{kang2025quits}, which we employ in our simulations.
Figure~\ref{fig:hgp_code_simulation_results} shows the results at $p=0.001$, where our cluster-based approach achieves approximately two orders of magnitude improvement in $\plog$ compared to the heuristic baselines, which is consistent with the performance gains observed for BB codes.
Additionally, the gap proxy with $N_\mathrm{class} = 18$ does not outperform the correction weight, again confirming its ineffectiveness.

\label{para:introducing_supplementary}
Additional simulation results are provided in the Supplementary Material:
\begin{itemize}
    \item Comprehensive data for all parameter combinations (Supplementary Figures~1--3).
    \item A detailed examination of the influence of the norm order~$\alpha$ (Supplementary Figure~4), showing that $\alpha = 1$ or $2$ is optimal for most cases.
    \item An evaluation of the alternative ``conv-max-conf'' setting, which treats BP-converged samples as having maximum decoding confidence (Supplementary Figure~5). 
    This setting avoids the need to modify BP+LSD to enforce LSD execution regardless of BP convergence, but is found to degrade post-selection performance in most cases.
    \item A comprehensive analysis of gap proxy heuristics and true logical gap (Supplementary Note~1, including Supplementary Table~1 and Supplementary Figures~6--8).
    We compare various heuristics for selecting which logical classes to decode, including uniform random sampling (used in Figs.~\ref{fig:bb_code_simulation_results} and~\ref{fig:hgp_code_simulation_results}), weighted sampling based on the empirical logical error distribution, and adaptive variants.
    Surprisingly, none of these heuristics outperform naive uniform random sampling.
    Through exhaustive comparative decoding across all $2^{12} = 4096$ logical classes for the $[[144, 12, 12]]$ BB code, we explain this counterintuitive result: post-selection performance depends on the \emph{discrimination power} between successful and failed samples, rather than how accurately the metric approximates the true logical gap.
    Distribution-based heuristics improve gap approximation primarily for successful shots but not for failed shots, thereby reducing discriminative power.
\end{itemize}

\subsubsection{Real-time strategy analysis}

We now analyze the real-time post-selection strategy.
To fairly assess the impact of ``aborting mid-way'', we evaluate the \textit{average time cost per accepted shot} $\avgTimeCostAcc$ instead of $\pabort$, by summing the total number of syndrome extraction rounds elapsed across all samples (including both aborted and accepted runs) and dividing by the number of accepted samples.
In other words, $\avgTimeCostAcc$ represents the expected time cost required to obtain a single accepted run when immediately retrying after each aborted run.
For comparison, global strategies yield $\avgTimeCostAcc = T / (1 - \pabort)$.

Our analysis focuses on $T$-round memory experiments using the surface code with distance $d = 13$ and the $[[144, 12, 12]]$ bivariate bicycle (BB) code.
We assume the same circuit-level noise model as the global strategy analysis with $p \in \{0.003, 0.005\}$.
The decoding is performed using the sliding window framework with window parameters $(W, F) = (5, 1)$ for the surface code and $(W, F) = (3, 1)$ for the BB code.
In both cases, the BP+LSD decoder serves as the inner decoder.
For post-selection, we employ the cluster LLR 2-norm fraction $\llrNormFrac{2}$ as our confidence metric with the lookback window size $L \in \{1, 2, 3, 5, 7\}$.
By varying the cutoff value for this metric, we obtain different trade-offs between $\plog$ and the average time cost per accepted shot $\avgTimeCostAcc$.

\begin{figure*}[!t]
	\centering
	\includegraphics[width=\textwidth]{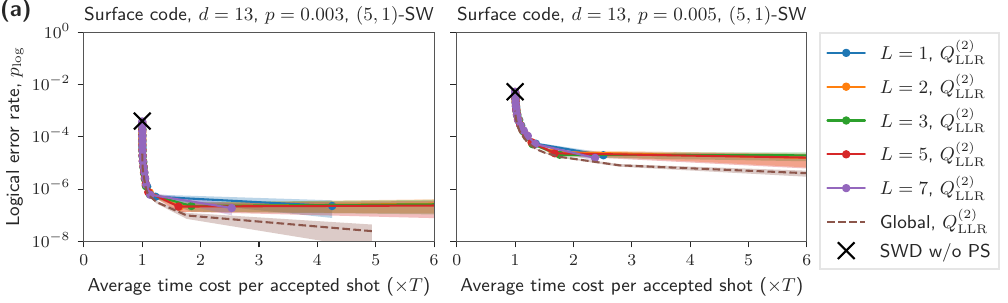}\\[0.5em]
    \includegraphics[width=\textwidth]{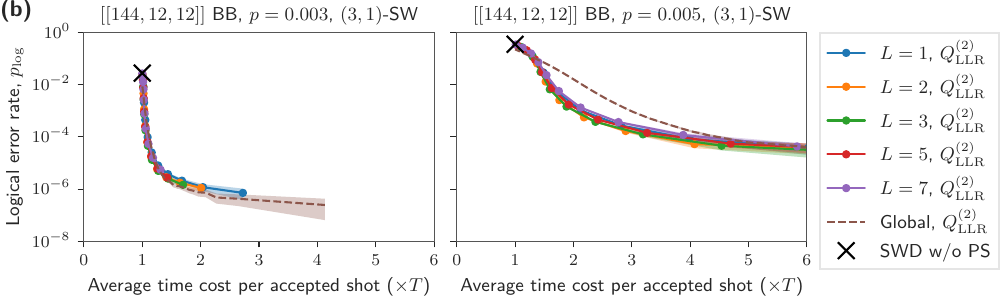}
	\caption{
        \textbf{Analysis of real-time post-selection via Strategy~\ref{strategy:realtime}} for \subfig{a} the surface code with $d=13$ and \subfig{b} the $[[144, 12, 12]]$ bivariate bicycle code.
        Logical error rates $\plog$ are plotted against the average time cost per accepted shots $\avgTimeCostAcc$, which represents the average time cost required to succeed when immediately retrying after each aborted attempt.
        The results are for the memory experiments with $T=d$ at $p \in \{0.003, 0.005\}$, decoded with the $(5, 1)$ or $(3, 1)$ sliding window method.
        The LLR 2-norm fraction $\llrNormFrac{2}$ is used for the metric and the parameter $L$ of the strategy varies across $\{1, 2, 3, 5, 7\}$.
        For comparison, the values for the global strategy (Strategy~\ref{strategy:global}) and those for sliding window decoding without post-selection are presented additionally as dashed lines and ``$\times$'' marks, respectively.
        Shaded regions represent 95\% confidence intervals.
    }
	\label{fig:real_time_post_selection_analysis}
\end{figure*}

Figure~\ref{fig:real_time_post_selection_analysis} presents our main results for both codes with $T=d$, clearly exhibiting the trade-off relation between $\plog$ and $\avgTimeCostAcc$.
For comparison, we include two baseline results: the global post-selection strategy (dashed lines) and standard sliding-window decoding without post-selection (``$\times$'' marks).
Notably, our real-time strategy achieves performance comparable to the global strategy in most cases, and specifically for the BB code at $p=0.005$, it even surpasses global post-selection performance by at most around 1.5 orders of magnitude in $\plog$.

\begin{figure*}[!t]
	\centering
    \includegraphics[width=\textwidth]{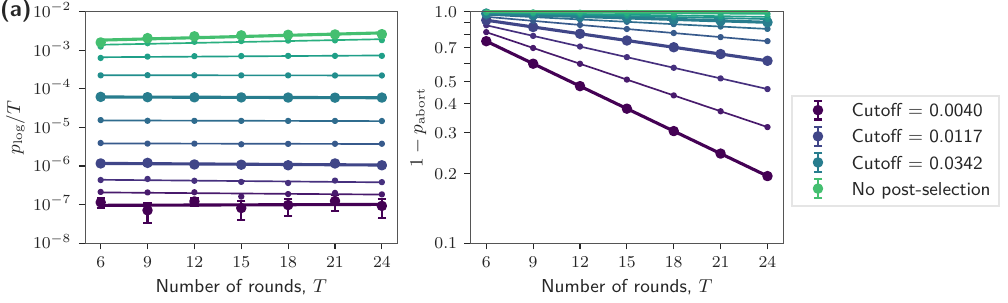}\\[0.5em]
    \includegraphics[width=0.49\textwidth]{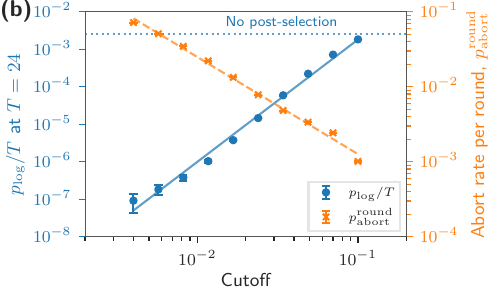}
    \includegraphics[width=0.4\textwidth]{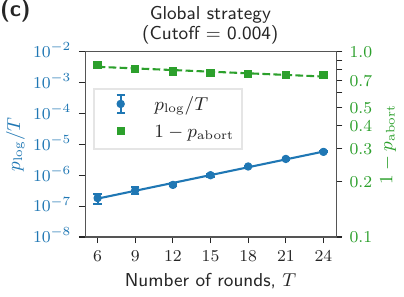}
	\caption{
        \textbf{Dependence of real-time post-selection performance on the number of rounds $T$} for the $[[144, 12, 12]]$ bivariate bicycle code.
        We consider the same setting as Fig.~\ref{fig:real_time_post_selection_analysis} at $p=0.003$ and $L = 3$ with varying $T \in \{6, 9, \cdots, 24\}$.
        All error bars represent 95\% confidence intervals.
        \subfig{a} The logical error rate per round ($\plog/T$) and acceptance rate ($1 - \pabort$) are plotted against $T$ for various cutoff values as well as the baseline scenario without post-selection.
        Three representative cutoff values (specified in the legend) and the ``no post-selection'' scenario are highlighted with bold lines and bigger markers.
        Both vertical axes are on a logarithmic scale, and the solid lines represent linear fits on this scale, showing that $\plog/T$ is almost constant while $\log(1 - \pabort)$ is nearly linear on $T$.
        \subfig{b} $\plog/T$ at the largest $T=24$ and the abort rate per round $\pabortround$ are plotted against the cutoff value.
        Here, $\pabortround \coloneqq 1 - 1/\exp(G)$, where $G$ is the slope of the linear fit of $\log(1 - \pabort)$ against $T$ in \subfig{a}.
        Solid and dashed lines represent linear fits on a log-log scale.
        Note that the linear fits are placed just for visual aid and may not be valid beyond the presented region.
        Specifically, the ``no post-selection'' scenario ($c = 1$) has $\plog/T \approx 2.6 \times 10^{-3}$ (blue dotted horizontal line) and $\pabortround = 0$, which deviate significantly from the lines.
        \subfig{c} Dependence of the global strategy performance on $T$ with the cutoff fixed to $0.004$ is presented for comparison.
        Unlike the real-time strategy, $\plog/T$ varies on $T$.
    }
	\label{fig:real_time_post_selection_over_rounds}
\end{figure*}

The real-time strategy can be particularly useful for circuits involving many syndrome extraction rounds, where the global strategy may become computationally prohibitive.
To assess the practical viability of our approach in such scenarios, we investigate how its performance scales with the total number of rounds $T$.
Figure~\ref{fig:real_time_post_selection_over_rounds}(a) examines the scaling behavior by plotting both $\plog/T$ and the acceptance rate $1 - \pabort$ (in a logarithmic scale) against $T$ for the $[[144, 12, 12]]$ BB code under various fixed cutoff values, including the scenario without post-selection as a baseline.
(Note that we here plot the acceptance rate rather than $\avgTimeCostAcc$, as it directly reveals its decay over rounds and allows us to estimate the per-round abort rate.)

Two key observations emerge from Fig.~\ref{fig:real_time_post_selection_over_rounds}(a):
First, $\plog/T$ remains nearly constant across different $T$ values even under post-selection, which indicates that logical errors accumulate linearly with the number of rounds, preserving the expected scaling behavior shown in the case without post-selection.
Second, $\log(1 - \pabort)$ exhibits clear linear dependence on $T$.
This behavior can be interpreted as every single round has an equal \textit{per-round abort rate} $\pabortround$, which is related to $\pabort$ as $\pabort = 1 - (1 - \pabortround)^T$ and thus can be estimated from the slope of the fit in the right panel.
In Fig.~\ref{fig:real_time_post_selection_over_rounds}(b), we plot the estimated values of $\plog/T$ (at $T = 24$) and $\pabortround$ against the cutoff, clearly illustrating the trade-off relation between them.
Together, these properties ensure that the real-time strategy offers predictable scaling behavior, allowing reliable extrapolation of performance to arbitrary number of rounds.

To highlight the advantages of the real-time approach, we conduct a parallel analysis for the global strategy in Fig.~\ref{fig:real_time_post_selection_over_rounds}(c) with a fixed cutoff of $c = 0.004$.
Crucially, $\plog/T$ is no longer invariant under $T$, underscoring the practical necessity of the real-time strategy particularly for large $T$.

\section{Discussion}

In this work, we have shown that decoder soft outputs derived from error-cluster structure provide an effective foundation for post-selection in quantum low-density parity check (QLDPC) codes. 
Leveraging the insight that clustering-based decoders (such as BP+LSD \cite{hillmann2025localized}) naturally capture the geometric properties of error configurations in the Tanner graph, we introduced two heuristic confidence metrics based on cluster statistics: cluster size and log-likelihood ratio (LLR) norm fractions.
Utilizing these metrics, we developed post-selection strategies that are applicable to general QLDPC codes and require only a single decoding run.
This approach offers significant advantages over the conventional logical gap method \cite{bombin2024fault,gidney2025yoked,smith2024mitigating}, which is effectively applicable only to a limited family of QEC codes (those compatible with the MWPM decoder, such as surface codes) and suffers from computational overheads that scale exponentially with the number of logical qubits.

Our numerical simulations demonstrated that our post-selection strategies achieve significant reductions in the logical error rate $\plog$ (for accepted runs) by several orders of magnitude at moderate abort rates $\pabort$. 
This performance gain is observed consistently across diverse code families: surface codes, bivariate bicycle codes, and hypergraph product codes (Figs.~\ref{fig:surface_code_simulation_results}--\ref{fig:hgp_code_simulation_results}).
For instance, by applying our strategy to the $[[144, 12, 12]]$ bivariate bicycle (BB) code, the logical error rate can be reduced by about three orders of magnitude with the abort rate of only around 1\% (19\%) at the physical error rate ($p$) of 0.1\% (0.3\%) under the uniform circuit-level noise model.
Compared to strategies based on other confidence metrics, the cluster-based approach significantly outperforms baseline metrics such as the correction weight and detector density across all code families.
While the logical gap method with MWPM remains superior for surface codes, cluster-based metrics outperform the logical gap from BP+LSD or its proxy calculated by partial comparative decoding with a subset of logical classes.

Our analysis highlights that post-selection can offer a far more cost-efficient alternative to simply increasing the code distance, particularly when non-determinism is acceptable and the system operates in a slightly sub-threshold regime. 
For example, at $p=0.1\%$ ($p=0.3\%$), aborting only about 8\% (4\%) of samples for the $[[72, 12, 6]]$ BB code achieves a logical error rate comparable to that of the larger $[[144, 12, 12]]$ BB code without post-selection. 
In other words, a modest spacetime cost increase of $\times 1.09$ ($\times 1.04$) with post-selection yields an effect roughly equivalent to a $\times 4$ increase in spacetime cost without post-selection.

Furthermore, real-time decoding is essential for practical quantum computing with non-Clifford gates, which require Clifford corrections conditioned on decoded measurement outcomes \cite{terhal2015quantum}.\label{para:realtime_discussion}
Such real-time decoding must be sufficiently fast to avoid the exponential backlog problem.
We extended our approach to this setting through sliding-window decoding, featuring mid-circuit abort decisions that terminate low-confidence trials early and save computational resources.
The real-time strategy achieves performance comparable to, or even superior to, global post-selection in terms of the average time cost per accepted shot $\avgTimeCostAcc$ for achieving a target logical error rate $\plog$ (Fig.~\ref{fig:real_time_post_selection_analysis}).
Notably, the real-time strategy exhibits favorable scaling with the number of rounds $T$, maintaining nearly constant per-round logical error and abort rates (Fig.~\ref{fig:real_time_post_selection_over_rounds}).
In contrast, the global strategy does not preserve this consistency; thus, the real-time approach offers more predictable and stable behavior for any $T$.

One may wonder whether real-time post-selection is truly necessary, given that for offline tasks (the main target of post-selection) backlog might not be the primary concern.
We believe it remains valuable, for several reasons:
(1) mid-circuit aborts can reduce expected resource costs by terminating low-confidence trials early, regardless of whether backlog is an issue;
(2) even offline tasks may require real-time decoding when they involve logical non-Clifford gates that necessitate intermediate Clifford corrections (e.g., multi-round magic state distillation, where logical non-Clifford gates are implemented by consuming magic states distilled from earlier rounds);
(3) decoder-confidence-based post-selection may also be useful for early fault-tolerant algorithms with moderate circuit depths, as suggested by recent studies on fault-tolerant error mitigation using decoder soft outputs~\cite{zhou2025error, dinca2025error}; and
(4) the favorable scaling with circuit depth demonstrated above can also benefit offline tasks with substantial depths, such as multi-round distillation.

We interpret the gains from cluster-based metrics as evidence that the geometry and concentration of errors (in terms of the connectivity structure on the Tanner graph) are key determinants of decoding ambiguity.
Large clusters are inherently problematic because they can span multiple local solutions and logical classes, thereby increasing the likelihood of miscorrection.
In contrast, fragmented, smaller clusters present fewer ambiguities and are more readily resolved by the decoder.
This geometric perspective explains why cluster norm fractions significantly outperform baseline metrics such as correction weight and detector density, which rely on simple aggregations of error weights or detector flips without capturing the detailed geometric structure of error configurations.
Additionally, the performance does not severely depend on the norm order $\alpha$ as long as $\alpha \geq 1$, which implies that, once large clusters dominate the metric, finer details of how residual clusters are formed matter little. 

Finally, we identify key limitations and promising directions for future work.\label{para:future_directions}
\begin{itemize}
    \item \textbf{Theoretical foundations:} Establishing theoretical justifications (e.g., bounds that relate cluster-based metrics to the logical gap) would provide rigorous foundations for these heuristics. Understanding which code or circuit properties (such as expansion \cite{leverrier2015quantum}) influence post-selection performance would also be valuable.
    \item \textbf{Further performance improvements:} While cluster-based metrics offer practical advantages, the logical gap method still outperforms them for surface codes.
    Developing better metrics that approach logical gap performance while retaining computational efficiency and generalizability remains an open challenge.
    One promising direction could be incorporating geometric information about clusters beyond their sizes, such as inter-cluster distances.
    However, inter-cluster distance alone would be insufficient: when only a single cluster exists, this distance is undefined, yet confidence can still vary significantly depending on the cluster's size.
    One might also consider distances between clusters and code boundaries, but this raises additional challenges for codes with periodic boundary conditions or general QLDPC codes where the notion of a boundary may not be rigorously defined.
    Designing optimal confidence metrics that appropriately incorporate such distance information remains a nontrivial open problem.
    \item \textbf{Gap proxy methods for multi-qubit codes:} Our gap proxy analysis (Supplementary Note~1) shows that various heuristics for selecting a subset of logical classes to decode do not outperform naive random sampling, and the resulting metric fails to outperform the correction weight method despite significant computational overhead.
    This is because post-selection performance depends on the discrimination power between successful and failed samples, not approximation accuracy; distribution-based heuristics improve accuracy primarily for successful shots, thereby reducing discriminative power.
    Effective gap proxy methods may require fundamentally different approaches.
    \item \textbf{Broader noise models:} Testing our strategies beyond standard circuit-level noise (including strongly biased, correlated, or other realistic noise models) would demonstrate their robustness. In these cases, cluster size and LLR norm fractions may lead to more pronounced differences in post-selection performance compared to our current settings.
    \item \textbf{Practical applications:} Evaluating the effectiveness of these strategies for specific quantum computing tasks, such as magic state preparation, would verify their practical utility.
    It would also be interesting to test whether these metrics are useful as decoder-confidence signals for logical-level error mitigation, analogous to recent studies using decoder soft information~\cite{zhou2025error,dinca2025error,aharonov2025syndrome}.
\end{itemize}

\section{Methods}

\subsection{Formal definitions of the graph-theoretic structures on clustering-based decoders}

In this subsection, we provide formal definitions of the graph-theoretic structures underlying clustering-based decoders, as introduced in the Results section.

For an $n$-dimensional vector $\vb{v}$, we denote its $i$-th entry by $v_i$ and, for an index set $I \subseteq \{1,\dots,n\}$, we write $\vb{v}[I]$ for the subvector consisting of $\{v_i : i \in I\}$, ordered by increasing indices.
For an $n \times m$ matrix $M$, we denote its $(i,j)$-th entry by $M[i,j]$ and, for $I_1 \subseteq \{1,\dots,n\}$ and $I_2 \subseteq \{1,\dots,m\}$, we write $M[I_1,I_2]$ for the submatrix with rows indexed by $I_1$ and columns indexed by $I_2$.

A detector error model (DEM) specifies a set of error mechanisms $\mathcal{E} = \{1, \ldots, N\}$, a set of detectors $\mathcal{D} = \{1, \ldots, r\}$, and the relationships between them.
The \textit{check matrix} $H \in \mathbb{Z}_2^{r \times N}$ is a binary matrix where $H[i, j] = 1$ if and only if detector $i$ is flipped by error mechanism $j$.
Each error mechanism $j$ has an associated prior probability $p_j \in [0, 1]$.

\begin{definition}[Tanner graph]
The \textit{Tanner graph} $\mathcal{G}_T = (\mathcal{D} \cup \mathcal{E}, E_T)$ is a bipartite graph where the vertex set is partitioned into check nodes (corresponding to detectors $\mathcal{D}$) and fault nodes (corresponding to error mechanisms $\mathcal{E}$).
An edge $(i, j) \in E_T$ exists between check node $i \in \mathcal{D}$ and fault node $j \in \mathcal{E}$ if and only if $H[i, j] = 1$.
The check matrix $H$ thus serves as the biadjacency matrix of the Tanner graph.
\label{def:tanner_graph}
\end{definition}

Given the Tanner graph structure, we can formally state the decoding problem.

\begin{definition}[Decoding problem]
Given a check matrix $H \in \mathbb{Z}_2^{r \times N}$ and a syndrome $\vb{s} \in \mathbb{Z}_2^r$ (a binary vector indicating which detectors are violated), the \textit{decoding problem} is to find a correction $\tilde{\vb{e}} \in \mathbb{Z}_2^N$ such that
\begin{align*}
    H \tilde{\vb{e}} = \vb{s}.
\end{align*}
In general, multiple solutions exist; a decoder selects one according to a specific strategy (e.g., minimizing the log-likelihood weight).
\label{def:decoding_problem}
\end{definition}

While the Tanner graph captures the relationship between error mechanisms and detectors, it is often useful to consider the connectivity structure among error mechanisms directly.
This motivates the construction of the fault graph.

\begin{definition}[Fault graph]
The \textit{fault graph} $\mathcal{G}_F = (\mathcal{E}, E_F)$ is constructed by projecting the Tanner graph onto the fault nodes.
Specifically, two fault nodes $j_1, j_2 \in \mathcal{E}$ (with $j_1 \neq j_2$) are connected by an edge in $E_F$ if and only if there exists at least one detector $i \in \mathcal{D}$ such that $H[i, j_1] = H[i, j_2] = 1$.
In other words, two error mechanisms are adjacent in the fault graph if they share a common detector.
The fault graph can equivalently be characterized through the Gram-like matrix $H^\top H$: an edge exists between $j_1$ and $j_2$ if and only if $(H^\top H)[j_1, j_2] \geq 1$.
\label{def:fault_graph}
\end{definition}

Clustering-based decoders operate by partitioning error mechanisms into groups that can be decoded independently.

\begin{definition}[Cluster]
A \textit{cluster} $C \subseteq \mathcal{E}$ is a set of error mechanisms that forms a connected subgraph in the fault graph $\mathcal{G}_F$.
Equivalently, for any two error mechanisms $j_1, j_2 \in C$, there exists a path in $\mathcal{G}_F$ connecting $j_1$ to $j_2$ that passes only through vertices in $C$.
\label{def:cluster}
\end{definition}

For a cluster to be useful for decoding, it must be ``self-contained'' in the sense that its internal structure suffices to determine a consistent correction.
This requirement is formalized through the notion of validity.

\begin{definition}[Valid cluster]
Let $C \subseteq \mathcal{E}$ be a cluster, and let $\mathcal{D}_C \coloneqq \{i \in \mathcal{D} : \exists j \in C \text{ such that } H[i, j] = 1\}$ denote the set of detectors that involve at least one error mechanism in $C$.
The cluster $C$ is \textit{valid} if there exists a correction $\tilde{\vb{e}} \in \mathbb{Z}_2^{N}$ such that for every detector $i \in \mathcal{D}_C$,
\begin{align*}
    \bigoplus_{j \in C} H[i, j] \cdot \tilde{e}_j = s_i,
\end{align*}
where $s_i \in \mathbb{Z}_2$ is the observed syndrome value for detector $i$, and ``$\oplus$'' denotes addition modulo 2.
In other words,
\begin{align}
    \vb{s}[\mathcal{D}_C] \in \image(H[\mathcal{D}_C, C]). \label{eq:cluster_validity} 
\end{align}
\label{def:valid_cluster}
\end{definition}

Intuitively, a valid cluster contains sufficient error mechanisms to explain all syndrome bits that it touches.
The validity condition ensures that the cluster can be decoded in isolation without affecting or being affected by error mechanisms outside the cluster.

We now formally define the class of decoders that leverage this cluster structure.

\begin{definition}[Clustering-based decoder]
A \textit{clustering-based decoder} is a decoder that solves the decoding problem (Definition~\ref{def:decoding_problem}) through the following procedure:
\begin{enumerate}
    \item \textit{Cluster construction}: Construct a collection of disjoint valid clusters $\{C_1, C_2, \ldots, C_m\}$ such that $\bigcup_{k=1}^m \mathcal{D}_{C_k}$ covers all violated detectors.
    \item \textit{Local decoding}: For each cluster $C_k$, solve the local decoding problem restricted to the submatrix $H[\mathcal{D}_{C_k}, C_k]$ to obtain a local correction $\tilde{\vb{e}}_k$.
    \item \textit{Correction aggregation}: Combine the local corrections to form the global correction $\tilde{\vb{e}}$, where $\tilde{e}_j = \tilde{e}_{k,j}$ if $j \in C_k$ for some $k$, and $\tilde{e}_j = 0$ otherwise.
\end{enumerate}
\label{def:clustering_based_decoder}
\end{definition}

The efficiency of clustering-based decoders stems from the fact that small, localized clusters require less computational effort to decode than solving the full decoding problem globally and can benefit from parallelization.

\subsection{Belief propagation plus localized statistics decoding (BP+LSD)}
\label{sec:methods_bplsd}

We briefly describe the BP+LSD decoder \cite{hillmann2025localized} using the notation established in the preceding subsection.
Given a check matrix $H \in \mathbb{Z}_2^{r \times N}$ and syndrome $\vb{s} \in \mathbb{Z}_2^r$, the decoder aims to find a correction $\tilde{\vb{e}} \in \mathbb{Z}_2^N$ satisfying $H \tilde{\vb{e}} = \vb{s}$.
BP+LSD first attempts to solve this problem using belief propagation (BP), and if BP fails to converge, it applies localized statistics decoding (LSD) as a postprocessing step.

\paragraph{Belief propagation phase.}
BP is executed on the Tanner graph $\mathcal{G}_T$ (Definition~\ref{def:tanner_graph}) for a fixed number of iterations.
This produces a posterior LLR $\ell_j \in \mathbb{R}$ for each error mechanism $j \in \mathcal{E}$, where smaller $\ell_j$ indicates higher posterior probability that the error mechanism occurred.
If BP converges to a valid solution $\tilde{\vb{e}}_\mathrm{BP}$ satisfying $H \tilde{\vb{e}}_\mathrm{BP} = \vb{s}$, this solution is returned.
Otherwise, the posterior LLRs $\{\ell_j\}_{j \in \mathcal{E}}$ are passed to the LSD phase as reliability guidance.

\paragraph{LSD cluster initialization, growth, and merging.}
LSD constructs valid clusters (Definition~\ref{def:valid_cluster}) by growing them from violated detectors.
For each detector $i$ with $s_i = 1$ (violated), an initially empty cluster $C$ is created with $\mathcal{D}_C = \{i\}$.
Each cluster is then grown incrementally by adding exactly one error mechanism per step.
To select which error mechanism to add, we define the boundary detectors $\beta(C) := \{i \in \mathcal{D}_C : \exists j \notin C \text{ with } H[i,j] = 1\}$ and the candidate error mechanisms $\Lambda(C) := \{j \notin C : \exists i \in \beta(C) \text{ with } H[i,j] = 1\}$.
At each step, the error mechanism with the smallest posterior LLR among candidates is added:
\begin{equation}
    j^\star := \arg\min_{j \in \Lambda(C)} \ell_j, \qquad C \leftarrow C \cup \{j^\star\}.
\end{equation}
When $j^\star$ connects to boundary detectors of multiple clusters, these clusters are merged by taking the union of their error mechanism sets.

\paragraph{Validity check.}
After each growth or merging step, the algorithm checks whether the updated cluster is valid, i.e., whether the local syndrome lies in the image of the local check matrix [Eq.~\eqref{eq:cluster_validity}].
This is implemented via \emph{on-the-fly elimination}: the decoder maintains an incremental elimination state (a row-echelon form over $\mathbb{Z}_2$) for $H[\mathcal{D}_C, C]$ and applies the same stored row operations to $\vb{s}[\mathcal{D}_C]$, yielding a reduced augmented system $[U \mid \vb{s}']$.
When the cluster grows by adding a new error mechanism, only the newly added column is eliminated using the stored operations (cluster merging is handled similarly).
The cluster is valid if and only if $[U \mid \vb{s}']$ contains no row of the form $[0 \cdots 0 \mid 1]$.

\paragraph{Local decoding and global assembly.}
Growth and merging continue until all clusters become valid.
For each valid cluster $C$, a local solution $\tilde{\vb{e}}_C$ satisfying $H[\mathcal{D}_C, C] \tilde{\vb{e}}_C = \vb{s}[\mathcal{D}_C]$ is computed using the maintained factorization.
The global correction $\tilde{\vb{e}}$ is assembled by placing each local solution on its corresponding coordinates and setting all other entries to zero.

\subsection{Sliding-window decoding}

We here describe the sliding-window decoding framework that we use for our real-time post-selection.
The framework with the window size $W$ and the commit size $F$ works as follows: 
The set of detectors is partitioned into a sequence of temporal windows, each spanning $W$ rounds, such that two consecutive windows overlap by $W - F$ rounds.
Within each window, a decoder identifies a correction; however, only a fraction of the correction associated with the first $F$ rounds of each window are committed to the global solution.
The syndrome is updated accordingly, and the decoder proceeds to the next window.
This process repeats until all detector rounds have been processed.

We now describe the algorithm more formally, using the notation introduced previously: the check matrix $H \in \mathbb{Z}_2^{r \times N}$, the sets of detectors $\mathcal{D}$ and error mechanisms $\mathcal{E}$, and the syndrome vector $\vb{s} \in \mathbb{Z}_2^r$.
Additionally, each detector $i \in \mathcal{D}$ is assigned a time coordinate $t_i \in \{0, 1, 2, \cdots \}$.
The algorithm maintains a global correction vector $\hat{\vb{e}}$, initially set to $\hat{\vb{e}} \leftarrow \vb{0} = (0, \cdots, 0) \in \mathbb{Z}_2^N$.

For each window $w \in \{0, 1, 2, \cdots\}$, the corresponding detector subset is defined as 
\begin{align*}
    \mathcal{D}_w \coloneqq \qty{i \in \mathcal{D} : wF \leq t_i \leq wF + W - 1 }.
\end{align*}
To construct the decoding sub-problem for window $w$, we identify relevant error mechanisms and detectors.
First, we determine the ``active error mechanisms'' for this window, which are error mechanisms that connect to detectors within the current window, excluding any that have already been committed in previous windows.
Formally, denoting the set of error mechanisms commited so far as $\mathcal{C}$ (initially empty), the set of active error mechanisms is given as
\begin{align*}
    \mathcal{E}_w \coloneqq \qty{j \in \mathcal{E} : \exists i \in \mathcal{D}_w \text{ s.t. } H[i,j] = 1} \setminus \mathcal{C},
\end{align*}
where $H[i,j]$ denotes the $(i,j)$ element of $H$.

The decoding sub-problem for window $w$ is then defined by the sub-matrix $H_w = H[\mathcal{D}_w, \mathcal{E}_w]$, the reduced syndrome vector $\vb{s}_w = \vb{s}[\mathcal{D}_w]$, and the reduced prior probability vector $\vb{p}_w = \vb{p}[\mathcal{E}_w]$.
This sub-problem can be solved using any chosen inner decoder.

After obtaining the window solution $\hat{\vb{e}}_w$ (satisfying $H_w \hat{\vb{e}}_w = \vb{s}_w$), only error mechanisms involved in detectors within the first $F$ rounds are committed: 
\begin{align}
    \mathcal{E}_w^{\text{commit}} \coloneqq \qty{j \in \mathcal{E}_w : \exists i \in \mathcal{D}_w \text{ s.t. } wF \leq t_i \leq wF + F - 1 \text{ and } H_{ij} = 1} \label{eq:committed_error_mechanisms}
\end{align}
Next, the global state is updated by incorporating the committed corrections.
To map the window-local solution back to the global problem, we define a commit vector $\hat{\vb{e}}_{w}^\mathrm{commit} \in \mathbb{Z}_2^N$ as
\begin{align*}
    \hat{\vb{e}}_{w}^\mathrm{commit}[j] = \begin{cases}
        \hat{\vb{e}}_w[\phi_w(j)] & \text{if } j \in \mathcal{E}_w^{\text{commit}}, \\
        0 & \text{otherwise},
    \end{cases}
\end{align*}
for each $j \in \mathcal{E}$, where $\phi_w$ maps global error mechanism indices to the corresponding local indices in $\mathcal{E}_w$.

The global correction vector and syndrome are then updated as $\hat{\vb{e}} \leftarrow \hat{\vb{e}} \oplus \hat{\vb{e}}_{w}^\mathrm{commit}$ and $\vb{s} \leftarrow \vb{s} \oplus H\hat{\vb{e}}_{w}^\mathrm{commit}$, where ``$\oplus$'' denotes addition modulo 2.
This update effectively removes the syndrome contributions from committed error corrections.
Finally, the set of committed error mechanisms is expanded: $\mathcal{C} \leftarrow \mathcal{C} \cup \mathcal{E}_w^{\text{commit}}$.

The algorithm continues until the windows decoded so far cover all detectors (i.e., $wF + W - 1 \geq \max \{t_i : i \in \mathcal{D} \}$).
In the final window $w_\mathrm{final}$, all remaining active error mechanisms are committed: $\mathcal{E}_{w_\mathrm{final}}^{\text{commit}} \coloneqq \mathcal{E}_{w_\mathrm{final}}$ instead of Eq.~\eqref{eq:committed_error_mechanisms}. 

\subsection{Technical details on the simulation methods \label{app:simulation_details}}

We here provide technical details on our numerical simulation methods used to obtain the data in the Results section.
 
\subsubsection{Circuit generation and sampling}
All simulations utilize the \texttt{stim} library \cite{gidney2021stim} to sample circuit measurement outcomes.
We generate the memory circuits for different QEC codes as follows:
\begin{itemize}
    \item \textbf{Surface codes:} Generated using \texttt{stim.Circuit.generated("surface\_code:rotated\_memory\_z")}.
    \item \textbf{BB codes:} Generated using the code provided in Ref.~\cite{gong2024toward}.
    \item \textbf{(3, 4)-regular HGP code:} Generated using the \texttt{QUITS} library \cite{kang2025quits} with the setting \texttt{HgpCode(h,h).build\_graph(seed=22)}, where \texttt{h} represents the following check matrix of the classical code that forms the HGP code:
    \begin{equation*}
        \begin{pmatrix}
        0 & 0 & 1 & 1 & 0 & 1 & 0 & 0 & 0 & 0 & 1 & 0\\
        0 & 1 & 0 & 1 & 0 & 0 & 1 & 0 & 0 & 0 & 0 & 1\\
        0 & 0 & 0 & 0 & 0 & 0 & 0 & 1 & 1 & 0 & 1 & 1\\
        0 & 0 & 0 & 0 & 1 & 0 & 1 & 0 & 0 & 1 & 1 & 0\\
        1 & 0 & 0 & 0 & 0 & 1 & 0 & 0 & 0 & 1 & 0 & 1\\
        0 & 1 & 1 & 0 & 0 & 0 & 0 & 0 & 1 & 1 & 0 & 0\\
        1 & 0 & 0 & 1 & 1 & 0 & 0 & 0 & 1 & 0 & 0 & 0\\
        0 & 1 & 0 & 0 & 1 & 1 & 0 & 1 & 0 & 0 & 0 & 0\\
        1 & 0 & 1 & 0 & 0 & 0 & 1 & 1 & 0 & 0 & 0 & 0
        \end{pmatrix}
    \end{equation*}
    This specific $[[225, 9, 6]]$ HGP code is identical to the one optimized in Ref.~\cite{kang2025quits}, featuring a depth-8 syndrome extraction circuit.
\end{itemize} 

\subsubsection{Noise model}
We apply a uniform circuit-level depolarizing noise model characterized by an error rate parameter $p$:
\begin{itemize}
    \item Single-qubit Clifford gates are followed by depolarizing errors with probability $p$ (i.e., each of $X$, $Y$, and $Z$ occurs with probability $p/3$).
    \item Two-qubit Clifford gates are followed by two-qubit depolarizing errors with probability $p$ (i.e., each of the 15 nontrivial two-qubit Pauli errors occurs with probability $p/15$).
    \item Idling qubits between consecutive time steps are subject to depolarizing errors with probability $p$.
    \item Resets in the $Z$ ($X$) basis are followed by an $X$ ($Z$) error with probability $p$.
    \item Measurement outcomes are randomly flipped with probability $p$.
\end{itemize}

\subsubsection{Decoder implementation}
While our post-selection strategies can be applied to any clustering-based decoder, we specifically employ the BP+LSD decoder \cite{hillmann2025localized} for our simulations,  implemented in the \texttt{ldpc} library \cite{roffee2022ldpc}.

The original decoder implementation skips LSD postprocessing when BP converges, which makes it impossible to obtain cluster-based metrics in converged cases.
To address this limitation, we modified the decoder to execute LSD regardless of BP convergence, and this modification has been integrated into the official \texttt{ldpc} library.
Additionally, we test our strategies under the alternative assumption that BP convergence indicates maximum confidence; these results are presented in the supplementary information.

We use the following parameters for BP+LSD across all simulations:
\begin{itemize}
    \item \texttt{max\_iter}: 30 (Same setting as in Ref.~\cite{hillmann2025localized})
    \item \texttt{bp\_method}: \texttt{"minimum\_sum"}
    \item \texttt{lsd\_method}: \texttt{"LSD\_0"}
\end{itemize}

\subsubsection{Overall pipeline implementation}
The complete simulation pipeline is implemented in our GitHub repository \texttt{ldpc-post-selection}\footnote{\url{https://github.com/seokhyung-lee/ldpc-post-selection}}.
The repository includes a \texttt{SoftOutputsBpLsdDecoder} class that supports executing the modified BP+LSD decoder (with optional sliding-window integration) and extracting both corrections and soft outputs.
Detailed usage instructions are provided in the repository documentation.

\section*{Data availability}

The aggregated data that support the findings of this study are openly available in Harvard Dataverse at \href{https://doi.org/10.7910/DVN/MNYLTA}{DOI:10.7910/DVN/MNYLTA}.
The raw data are not uploaded due to their size but are available upon reasonable request to the authors.

\section*{Code availability}

The source codes used for the simulations in this work, including the \texttt{ldpc-post-selection} package, are available in GitHub: \url{https://github.com/seokhyung-lee/ldpc-post-selection}.

\bibliographystyle{quantum}
\bibliography{references}

\section*{Acknowledgements}

We thank Timo Hillmann, Nicholas Fazio, Dominic Williamson, Mingyu Kang, and Hyukgun Kwon for helpful discussions and comments.
We also thank Joschka Roffe for integrating our BP+LSD customization into the official \texttt{ldpc} library.

\section*{Funding}

This work is supported by the Australian Research Council via the Centre of Excellence in Engineered Quantum Systems (EQUS) Project No. CE170100009, and by the Intelligence Advanced Research Projects Activity (IARPA) through the Entangled Logical Qubits program Cooperative Agreement Number W911NF-23-2-0223.

\section*{Author contributions}

S.H.L. conceived the idea and all authors contributed to developing it. S.H.L. conducted simulations and all authors interpreted the results together. S.H.L. mainly wrote the manuscript, and L.E. and S.B. contributed helpful feedback and revisions. All authors reviewed the manuscript.

\section*{Competing interests}

The authors declare no competing interests.

\end{document}

\title{Supplementary Information for ``Efficient Post-Selection for General Quantum LDPC Codes''}

\author{Seok-Hyung Lee}
\affiliation{Centre for Engineered Quantum Systems, School of Physics, The University of Sydney, Sydney, New South Wales 2006, Australia}
\affiliation{Department of Quantum Information Engineering, Sungkyunkwan University, Suwon 16419, Republic of Korea}

\author{Lucas H. English}
\affiliation{Centre for Engineered Quantum Systems, School of Physics, The University of Sydney, Sydney, New South Wales 2006, Australia}

\author{Stephen D. Bartlett}
\affiliation{Centre for Engineered Quantum Systems, School of Physics, The University of Sydney, Sydney, New South Wales 2006, Australia}

\maketitle

\begin{figure*}[!h]
	\centering
	\includegraphics[width=\textwidth]{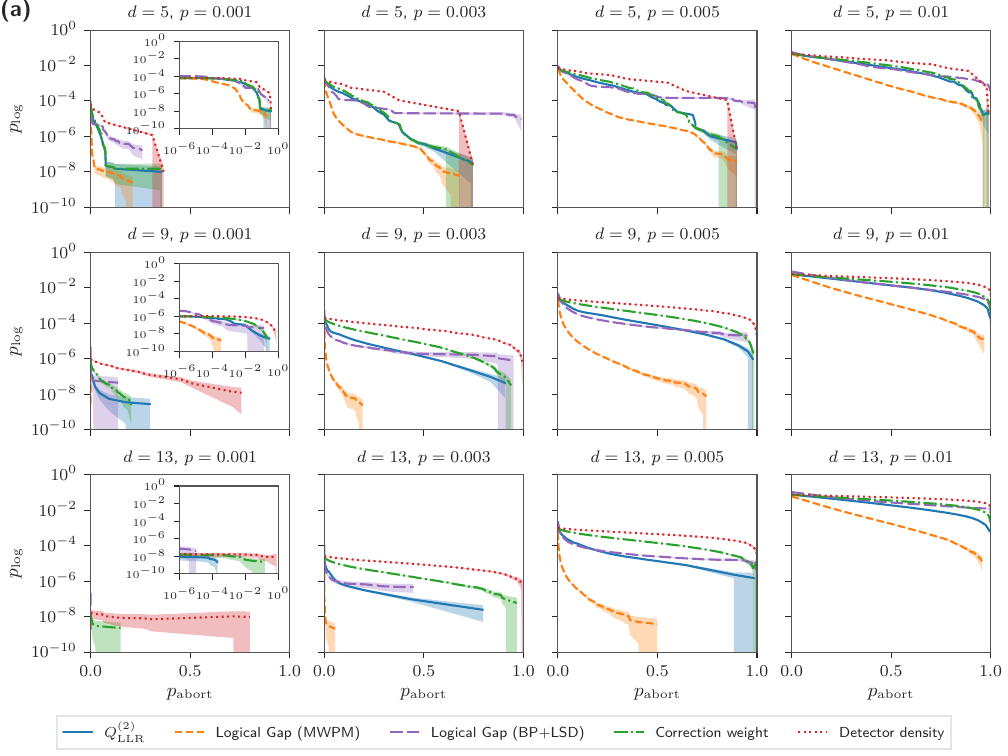}\\[0.5em] 
    \includegraphics[width=\textwidth]{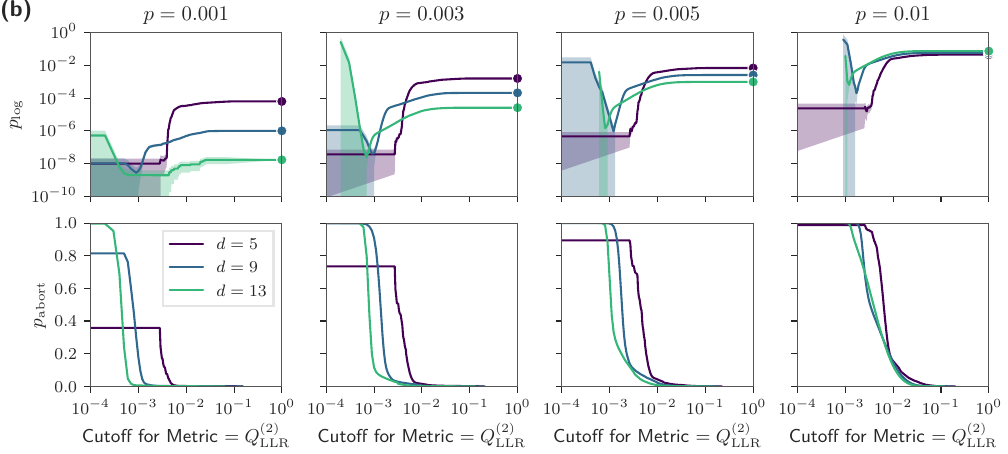}
	\caption{
        \textbf{Global post-selection analysis with surface codes for all parameter combinations.}
        All shaded regions represent 95\% confidence intervals.
        \subfig{a} Logical error rates $\plog$ are plotted against the abort rates $\pabort$ for five decoding confidence metrics: the cluster LLR 2-norm fraction $\llrNormFrac{2}$, the logical gap (MWPM), the logical gap (BP+LSD), correction weight, and detector density.
        This is an extension of Fig.~\ref{fig:surface_code_simulation_results}(b) in the main text.
        \subfig{b} $\plog$ and $\pabort$ are separately plotted against the cutoff values for the metric $\llrNormFrac{2}$.
    }
	\label{fig:surface_code_analysis_full}
\end{figure*}

\begin{figure*}[!h]
	\centering
	\includegraphics[width=0.8\textwidth]{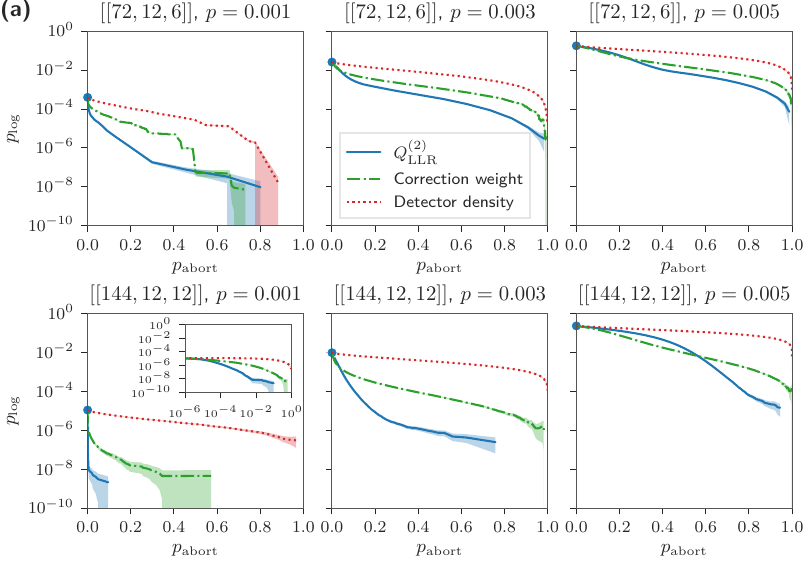}\\[0.5em]
    \includegraphics[width=0.8\textwidth]{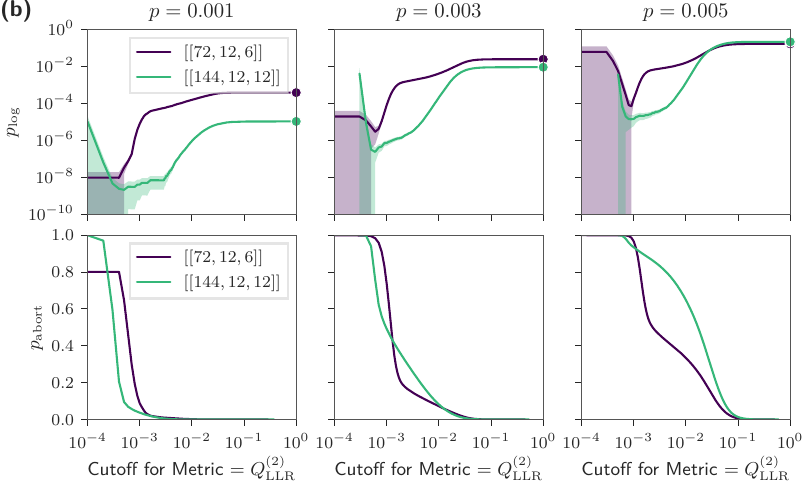}
	\caption{
        \textbf{Global post-selection analysis with bivariate bicycle codes for all parameter combinations.}
        All shaded regions represent 95\% confidence intervals.
        \subfig{a} Any-observable logical error rates $\plog$ are plotted against the abort rates $\pabort$ for three decoding confidence metrics: the cluster LLR 2-norm fraction $\llrNormFrac{2}$, correction weight, and detector density.
        This is an extension of Fig.~\ref{fig:bb_code_simulation_results}(b) in the main text.
        \subfig{b} $\plog$ and $\pabort$ are separately plotted against the cutoff values for the metric $\llrNormFrac{2}$.
    }
    
	\label{fig:bb_code_plog_vs_pabort_full}
\end{figure*}

\begin{figure*}[!h]
	\centering
	\includegraphics[width=0.5\textwidth]{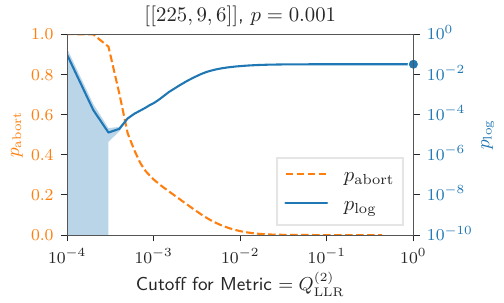}
	\caption{
        \textbf{Global post-selection analysis with the $[[225, 9, 6]]$ hypergraph product code.}
        $\plog$ and $\pabort$ are plotted against the cutoff values for the metric $\llrNormFrac{2}$ at $p=0.001$.
    }
	\label{fig:hgp_code_performance_vs_metric_full}
\end{figure*}

\begin{figure*}[!h]
	\centering
	\includegraphics[width=\textwidth]{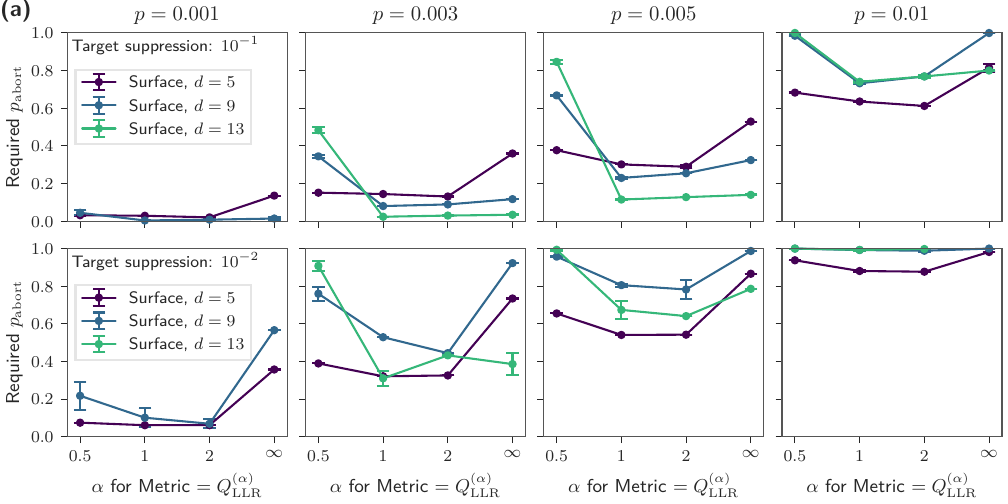}\\[0.5em]
    \includegraphics[width=0.7\textwidth]{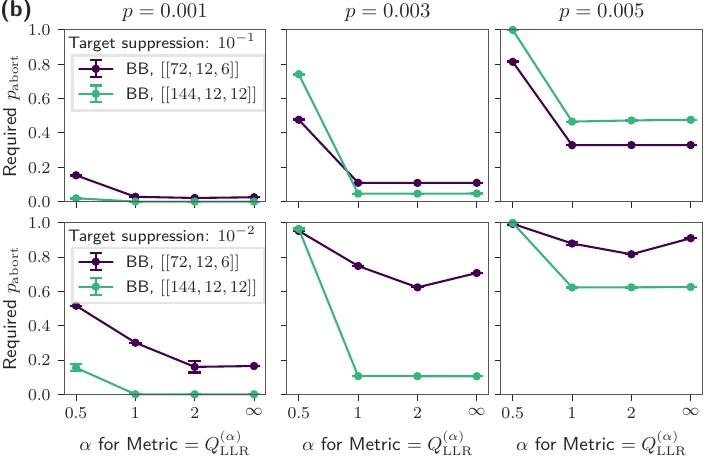}
    \includegraphics[width=0.29\textwidth]{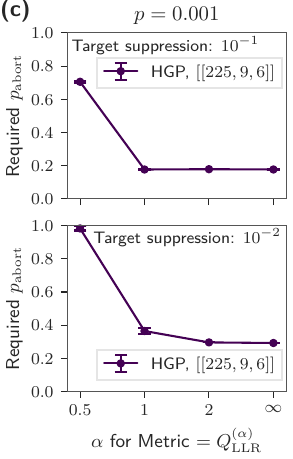}
	\caption{
        \textbf{Dependence of global post-selection performance on the norm order $\alpha$.}
        Abort rates $\pabort$ required to achieve logical error rate suppression of $10^{-1}$ or $10^{-2}$ (relative to cases without post-selection) are presented for different $\alpha$ values in cluster LLR norm fractions $\llrNormFrac{\alpha}$. 
        \subfig{a}, \subfig{b}, and \subfig{c} are for surface, BB, and HGP codes, respectively. 
        All error bars represent 95\% confidence intervals and solid lines are drawn just for visual aid.
        Overall, $\alpha=1$ and $\alpha=2$ show the best results, while $\alpha=0.5$ and $\alpha=\infty$ have degraded performance.
    }
	\label{fig:alpha_comparison}
\end{figure*}

\begin{figure*}[!h]
	\centering
	\includegraphics[width=0.49\textwidth]{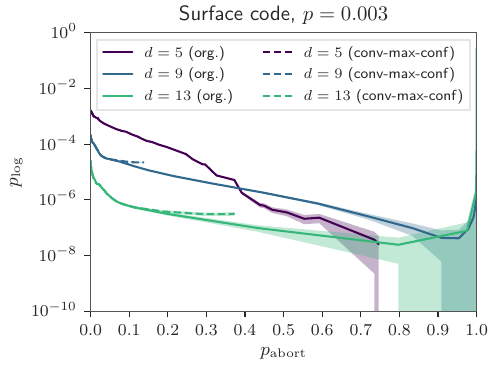}
    \includegraphics[width=0.49\textwidth]{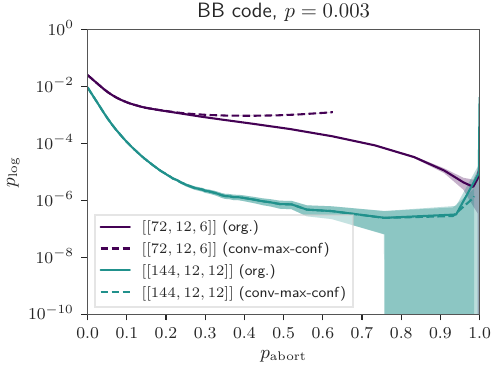}
	\caption{
        \textbf{Global post-selection analysis assuming BP convergence as maximum decoding confidence.}
        Our simulations employ the BP+LSD decoder modified to execute LSD even if BP converges, which is necessary to ensure cluster-based metrics are always obtainable.
        These plots analyze an alternative approach: assuming converged cases to have maximum decoding confidence (which makes this modification of the decoder unnecessary).
        $\plog$ is plotted against $\pabort$ for this approach as dashed lines (labeled as ``conv-max-conf'') for surface and BB codes at $p=0.003$.
        Since samples with converged BP are always accepted, $\pabort$ is upper bounded by $1 - \text{(convergence rate)}$.
        Compared to the original results (solid lines), this alternative approach has degraded performance in most cases, where achievable $\plog$ is lower-bounded by the logical error rate conditioned on convergence.
    }
	\label{fig:bp_convergence_analysis} 
\end{figure*}

\clearpage

\section{Gap proxy and exhaustive comparative decoding analysis}

As discussed in the main text, the logical gap method faces computational challenges for codes with multiple logical qubits, requiring comparative decoding across $2^l$ logical classes for $l$ nontrivial independent logical Pauli errors (where $l \leq 2k$ for $k$ logical qubits).
For instance, the $[[144, 12, 12]]$ bivariate bicycle (BB) code has $2^{12} = 4096$ logical classes when decoding with respect to only logical $Z$ observables.
To address this computational barrier, we introduce the \emph{gap proxy}, which approximates the logical gap by comparative decoding across a subset of $N_\mathrm{class} < 2^l$ classes composed of the standard BP+LSD correction and $N_\mathrm{class} - 1$ additional classes.

We test five heuristics for selecting the additional classes, defined in Supplementary Table~\ref{tab:gap_proxy_heuristics}.
All the heuristics except ``\texttt{random}'' utilize a pre-computed empirical logical error distribution, estimated from a sufficient number of shots of BP+LSD decoding without post-selection.
The underlying intuition is that frequently occurring logical errors are more likely to correspond to competitive alternative classes in the logical gap calculation; therefore, prioritizing these frequent errors might improve the gap proxy's accuracy compared to random sampling.

\begin{table*}[!b]
    \centering
    \caption{
        \textbf{Definitions of gap proxy heuristics.}
        The gap proxy approximates the logical gap by comparative decoding across a subset of $N_\mathrm{class}\leq 2^l$ classes, composed of the standard BP+LSD correction and $N_\mathrm{class} - 1$ additional classes selected in a specific way determined by the heuristic.
    }
    \label{tab:gap_proxy_heuristics}
    \begin{tabular}{@{}lp{9.3cm}@{}}
        \toprule
        Heuristic & Definition \\
        \midrule
        \texttt{random} & Uniformly sample $N_\mathrm{class} - 1$ classes from the remaining $2^l - 1$ classes. \\[0.5em]
        \texttt{most-likely-first} & Select the $N_\mathrm{class} - 1$ most probable logical errors from the empirical distribution, then obtain the corresponding classes by applying these errors to the correction from standard decoding. \\[0.5em]
        \texttt{weighted-random} & Sample $N_\mathrm{class} - 1$ logical errors with probabilities proportional to their frequencies in the logical error distribution, then obtain the corresponding classes. \\[0.5em]
        \texttt{adaptive-most-likely-first} & Proceed as \texttt{most-likely-first} by decoding classes one by one, but if a class with lower correction weight is found, update the base class for applying the logical error distribution to that class and continue decoding from the most-likely class (skipping already explored classes). \\[0.5em]
        \texttt{adaptive-weighted-random} & Proceed as \texttt{weighted-random} by decoding classes one by one, but if a class with lower correction weight is found, update the base class for applying the logical error distribution and continue decoding. \\
        \bottomrule
    \end{tabular}
\end{table*}

\begin{figure*}[!t]
	\centering
	\includegraphics[width=\textwidth]{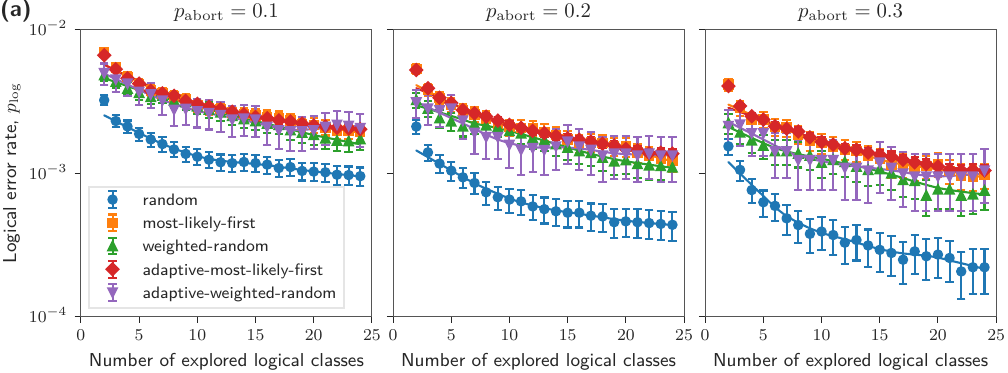}\\[0.5em]
    \includegraphics[width=0.49\textwidth]{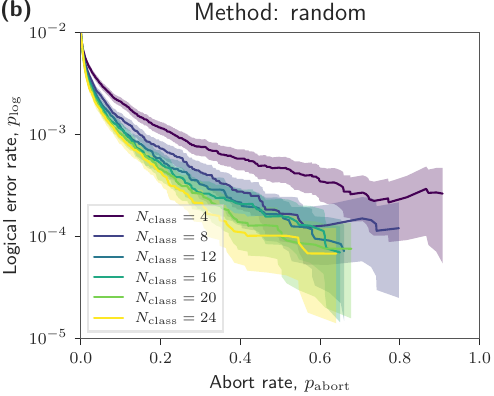}
    \includegraphics[width=0.49\textwidth]{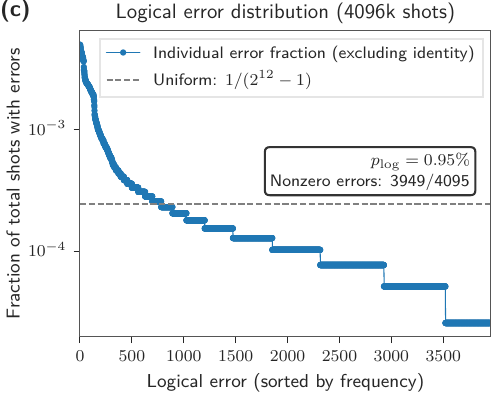}
	\caption{
        \textbf{Gap proxy analysis for the $[[144, 12, 12]]$ bivariate bicycle code at $p = 0.003$.}
        All error bars/bands represent 95\% confidence intervals.
        \subfig{a} Logical error rates $\plog$ are plotted against $N_{\text{class}}$ for different heuristics (see Supplementary Table~\ref{tab:gap_proxy_heuristics}) at fixed abort rates $\pabort \in \{0.1, 0.2, 0.3\}$.
        Lines are LOWESS fits with smoothing parameter 0.4.
        \subfig{b} Trade-off between $\plog$ and $\pabort$ for the \texttt{random} heuristic at various $N_{\text{class}}$.
        \subfig{c} Empirical logical error distribution from \num{4096000} shots (sorted by frequency, renormalized excluding the identity which accounts for $\sim$95\% of shots).
    }
	\label{fig:bb_gap_proxy_analysis}
\end{figure*}

\paragraph*{Gap proxy analysis for the BB code.}
Supplementary Figure~\ref{fig:bb_gap_proxy_analysis}(a) compares these heuristics for the $[[144, 12, 12]]$ BB code at $p = 0.003$.
Surprisingly, uniform random sampling (\texttt{random}) performs best across all settings, contrary to the intuition that distribution-based heuristics should help.
Supplementary Figure~\ref{fig:bb_gap_proxy_analysis}(b) shows the trade-off between $\plog$ and $\pabort$ for the \texttt{random} heuristic at various $N_{\text{class}}$, and Supplementary Figure~\ref{fig:bb_gap_proxy_analysis}(c) presents the empirical logical error distribution from \num{4096000} shots.
The distribution is highly non-uniform, with a small number of logical errors dominating, yet heuristics that prioritize these frequent errors do not improve gap proxy performance.

\begin{figure*}[!t]
	\centering
	\includegraphics[width=\textwidth]{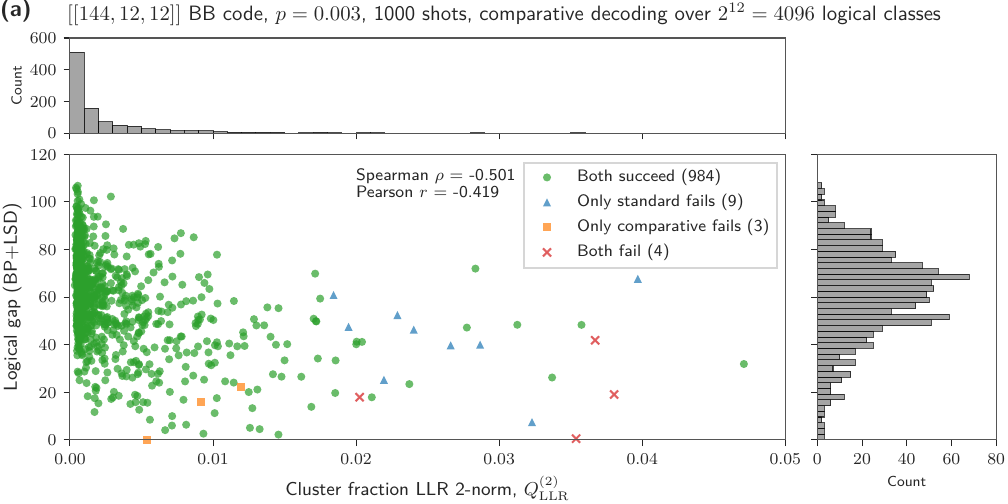}\\[0.5em]
    \includegraphics[width=\textwidth]{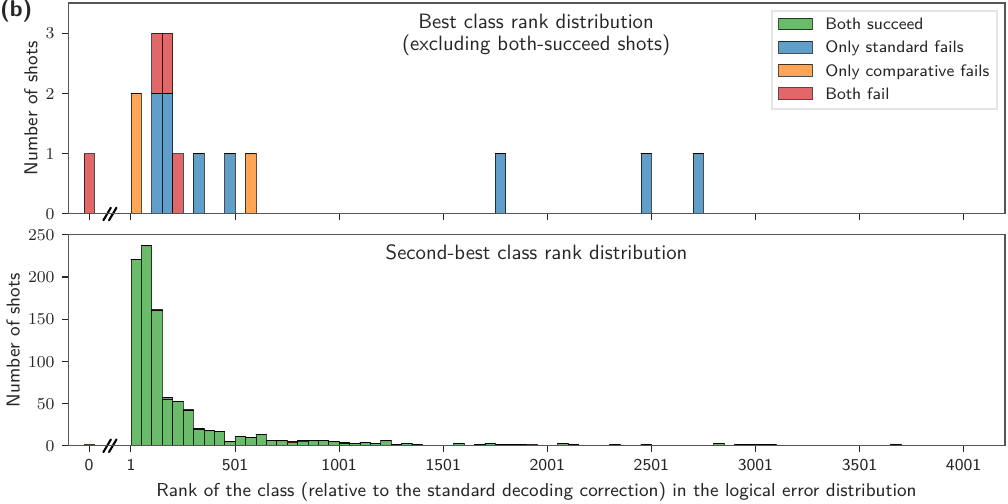}
	\caption{
        \textbf{Exhaustive comparative decoding analysis for the $[[144, 12, 12]]$ bivariate bicycle code.}
        We perform comparative decoding with BP+LSD across all $2^{12} = 4096$ logical classes for 1000 shots at $p = 0.003$.
        \subfig{a} 2D scatter plot with marginal histograms showing the correlation between $\llrNormFrac{2}$ and the logical gap.
        Points are colored by decoding outcome: ``Both succeed'' (green circles, 984 shots); ``Only standard fails'' (blue triangles, 9 shots); ``Only comparative fails'' (orange squares, 3 shots); ``Both fail'' (red crosses, 4 shots).
        \subfig{b} Rank distributions (in the empirical logical error distribution) of $\va{\lambda}_\mathrm{std} \oplus \va{\lambda}_\mathrm{(1)}$ and $\va{\lambda}_\mathrm{std} \oplus \va{\lambda}_\mathrm{(2)}$, where $\va{\lambda}_\mathrm{std}$ is the logical flip vector from standard decoding and $\va{\lambda}_\mathrm{(1)/(2)}$ corresponds to the best/second-best class from comparative decoding.
        Cases where standard decoding is the best class (98.4\% of samples) are omitted from the top panel.
    }
	\label{fig:exhaustive_gap_analysis}
\end{figure*}

\begin{figure*}[!t]
	\centering
	\includegraphics[width=\textwidth]{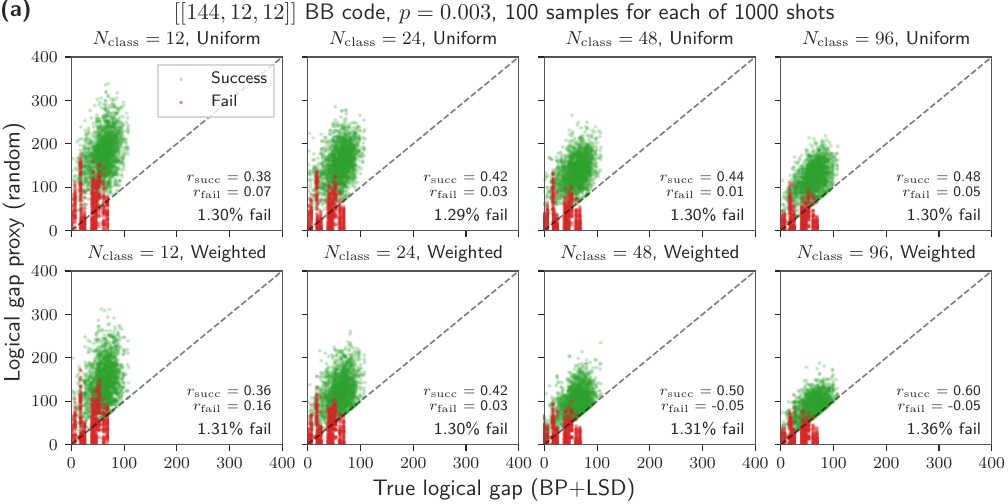}\\[0.5em]
    \includegraphics[width=\textwidth]{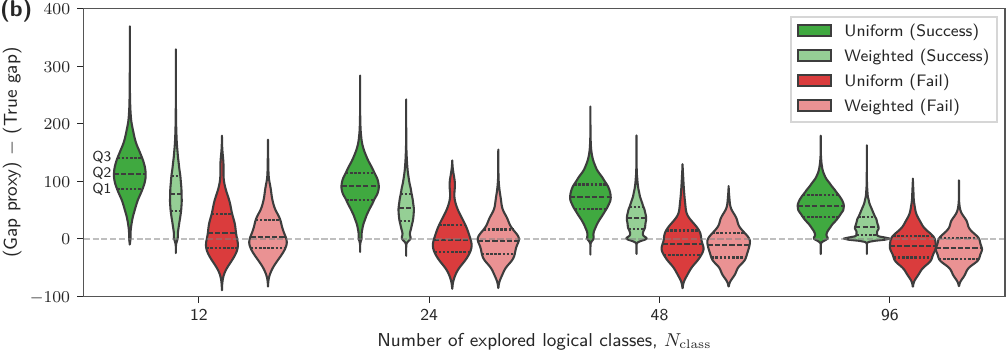}
	\caption{
        \textbf{Sampling-based analysis of \texttt{random} and \texttt{weighted-random} gap proxy heuristics for the $[[144, 12, 12]]$ bivariate bicycle code.}
        For each of 1000 shots from the exhaustive analysis (Supplementary Figure~\ref{fig:exhaustive_gap_analysis}), we sample 100 gap proxy values using either \texttt{random} (``Uniform'') or \texttt{weighted-random} (``Weighted'') heuristics at $N_{\text{class}} \in \{12, 24, 48, 96\}$.
        \subfig{a} Scatter plots of gap proxy versus true logical gap.
        Green (red) points indicate successful (failed) samples.
        Pearson correlation coefficients $r_{\text{succ}}$ and $r_{\text{fail}}$ are shown for each condition.
        \subfig{b} Distributions of $\text{(gap proxy)} - \text{(true gap)}$ for uniform and weighted sampling, separated by decoding success/failure.
    }
	\label{fig:gap_proxy_distribution_analysis}
\end{figure*}

\paragraph*{Exhaustive comparative decoding analysis.}
To understand why distribution-based heuristics fail to improve performance, we perform exhaustive comparative decoding with BP+LSD across all $2^{12} = 4096$ logical classes for 1000 shots at $p = 0.003$ (Supplementary Figures~\ref{fig:exhaustive_gap_analysis} and~\ref{fig:gap_proxy_distribution_analysis}).
Supplementary Figure~\ref{fig:exhaustive_gap_analysis}(a) shows the correlation between the cluster LLR 2-norm fraction $\llrNormFrac{2}$ and the true logical gap.
The negative correlation (Spearman $\rho = -0.501$, Pearson $r = -0.419$) suggests a modest tendency for larger cluster fractions to correspond to lower logical gaps, consistent with both metrics reflecting decoding confidence.

Supplementary Figure~\ref{fig:exhaustive_gap_analysis}(b) shows the rank distributions (in the empirical logical error distribution) of $\va{\lambda}_\mathrm{std} \oplus \va{\lambda}_\mathrm{(1)}$ and $\va{\lambda}_\mathrm{std} \oplus \va{\lambda}_\mathrm{(2)}$, where $\va{\lambda}_\mathrm{std}$ is the logical flip vector corresponding to the standard decoding correction and $\va{\lambda}_\mathrm{(1)/(2)}$ corresponds to the best/second-best class from comparative decoding.
The second-best class tends to have relatively low ranks (e.g., 836 out of 1000 shots have rank $\leq 500$), which supports the intuition behind distribution-based heuristics.

\paragraph*{Why distribution-based heuristics fail.}
However, as shown in Supplementary Figure~\ref{fig:bb_gap_proxy_analysis}(a), this intuition does not translate to improved post-selection performance.
To investigate the origin of this counterintuitive result, for each of 1000 shots from the exhaustive analysis, we sample 100 gap proxy values using either uniform (\texttt{random}) or weighted (\texttt{weighted-random}) sampling at various $N_{\text{class}}$ values.
The resulting distributions of the true logical gap and its proxy are presented in Supplementary Figure~\ref{fig:gap_proxy_distribution_analysis}(a), which provides two observations:
\begin{enumerate}
    \item Successful and failed shots exhibit markedly different distributions; successful shots show moderate correlation ($r_\text{succ} \in [0.36, 0.60]$) with points spread above the $y = x$ line, while failed shots show little correlation ($r_\text{fail} \in [-0.05, 0.16]$).
    \item Switching from uniform to weighted sampling moves successful shots closer to the $y = x$ line, but the distribution of failed shots remains largely unchanged.
\end{enumerate}

Supplementary Figure~\ref{fig:gap_proxy_distribution_analysis}(b) confirms this asymmetry: Weighted sampling shifts the success distribution toward zero (reducing overestimation), while the fail distribution barely changes.
A key insight here is that post-selection performance depends on the \emph{discrimination power} between successful and failed samples, rather than how accurately the metric approximates the true logical gap.
Thus, if a class selection strategy enhances approximation quality predominantly for successful shots, it can ironically degrade post-selection performance by making the metric distributions for successful and failed shots more similar.

The root cause of this asymmetry can be understood as follows.
Our approach first runs standard decoding once to obtain the class $\lambda_\mathrm{std}$ corresponding to the correction, then selects the remaining $N_\mathrm{class} - 1$ classes according to some strategy.
\begin{itemize}
    \item When standard decoding succeeds, the correction from $\lambda_\mathrm{std}$ almost always has the minimum weight across all classes; see Supplementary Figure~\ref{fig:exhaustive_gap_analysis}(b, top panel) where the best class rank is nonzero for only three shots (corresponding to `only comparative fails') out of 987 `standard succeeds' shots.
    Therefore, to compute the correct logical gap, it suffices to find the second-best class among the sampled classes in most cases.
    \item When standard decoding fails, the correction $\lambda_\mathrm{std}$ may not correspond to the best class; see Supplementary Figure~\ref{fig:exhaustive_gap_analysis}(b, top panel) where the best class rank is nonzero for 12 shots out of 13 `standard fails' shots.
    In such cases, computing the correct logical gap requires finding \emph{both} the best and second-best classes from the sampled subset, which is inherently more difficult.
\end{itemize}
Consequently, any class selection strategy is likely to be more effective at improving gap approximation for successful shots than for failed shots.
Moreover, distribution-based heuristics apply the empirical logical error distribution relative to $\lambda_\mathrm{std}$, which may differ from the true best class when decoding fails; this mismatch causes the heuristics to mis-rank candidate classes for failed shots, further limiting their effectiveness.
(The `adaptive' variants may mitigate this issue by updating the base class whenever a better class is found, but its effectiveness is still limited when $N_\mathrm{class}$ is small.)
This fundamental asymmetry explains why distribution-based heuristics, despite their intuitive appeal, do not outperform naive random sampling: They primarily improve approximation accuracy for successful shots, thereby reducing the metric's discriminative power for post-selection.